\documentclass[12pt,preprint]{aastex}

\usepackage{amsmath}
\usepackage{graphicx}
\usepackage{multirow}
\usepackage{color}
\usepackage{natbib}
\usepackage{url}

\begin{document} 

\title{OSSOS III -- Resonant Trans-Neptunian Populations: Constraints from the first quarter of the Outer Solar System Origins Survey}

\author{Kathryn Volk\altaffilmark{1}, Ruth Murray-Clay\altaffilmark{2}, Brett Gladman\altaffilmark{3}, Samantha Lawler\altaffilmark{4}, Michele T. Bannister\altaffilmark{4,5}, J. J. Kavelaars\altaffilmark{4,5}, Jean-Marc Petit\altaffilmark{6}, Stephen Gwyn\altaffilmark{4}, Mike Alexandersen\altaffilmark{7},  Ying-Tung Chen\altaffilmark{7}, Patryk Sofia Lykawka\altaffilmark{8}, Wing Ip\altaffilmark{9}, Hsing Wen Lin\altaffilmark{9}}

\altaffiltext{1}{Department of Planetary Sciences/Lunar and Planetary Laboratory, University of Arizona, 1629 E University Blvd, Tucson, AZ 85721}
\altaffiltext{2}{Department of Physics, University of California Santa Barbara}
\altaffiltext{3}{Department of Physics and Astronomy, University of British Columbia, 6224 Agricultural Road, Vancouver, BC V6T 1Z1, Canada}
\altaffiltext{4}{NRC, National Research Council of Canada, 5071 West Saanich Rd, Victoria, British Columbia V9E 2E7, Canada}
\altaffiltext{5}{Department of Physics and Astronomy, University of Victoria, Elliott Building, 3800 Finnerty Rd, Victoria, British Columbia V8P 5C2, Canada}
\altaffiltext{6}{Observatoire de Besancon, Universite de Franche Comte -- CNRS, Institut UTINAM, 41
bis avenue de l'Observatoire, 25010 Besancon Cedex, France}
\altaffiltext{7}{Institute for Astronomy and Astrophysics, Academia Sinica, Taiwan}
\altaffiltext{8}{Astronomy Group,  School of Interdisciplinary Social and Human Sciences, Kinki University, Japan}
\altaffiltext{9}{Institute of Astronomy, National Central University, Taiwan}

\begin{abstract}
The first two observational sky ``blocks" of the Outer Solar System Origins Survey (OSSOS) have significantly increased the number of well-characterized observed trans-Neptunian objects (TNOs) in Neptune's mean motion resonances.  We describe the 31 securely resonant TNOs detected by OSSOS so far, and we use them to independently verify the resonant population models from the Canada-France Ecliptic Plane Survey \citep[CFEPS;][]{Gladman2012}, with which we find broad agreement.   We confirm that the 5:2 resonance is more populated than models of the outer Solar System's dynamical history predict; our minimum population estimate shows that the high eccentricity ($e>0.35$) portion of the resonance is at least as populous as the 2:1 and possibly as populated as the 3:2 resonance. One OSSOS block was well-suited to detecting objects trapped at low libration amplitudes in Neptune's 3:2 resonance, a population of interest in testing the origins of resonant TNOs.  We detected three 3:2 objects with libration amplitudes below the cutoff modeled by CFEPS; OSSOS thus offers new constraints on this distribution.  The OSSOS detections confirm that the 2:1 resonance has a dynamically colder inclination distribution than either the 3:2 or 5:2 resonances.  Using the combined OSSOS and CFEPS 2:1 detections, we constrain the fraction of 2:1 objects in the symmetric mode of libration to be 0.2--0.85; we also constrain the fraction of leading vs. trailing asymmetric librators, which has been theoretically predicted to vary depending on Neptune's migration history, to be 0.05--0.8. Future OSSOS blocks will improve these constraints. 

\end{abstract}

\section{Introduction}\label{s:intro}

Trans-Neptunian objects (TNOs) are a dynamically diverse population of minor planets in the outer Solar System. A striking feature of the observed TNOs is the significant number of objects found in mean motion resonance with Neptune. Neptune's population of primordially captured resonant objects provides an important constraint on Solar System formation and giant planet migration scenarios \citep[e.g.,][]{Malhotra1995,Chiang2002,Hahn2005,Murray-Clay2005,Levison2008,Morbidelli2008,Nesvorny2015}.  But to understand these constraints on the early Solar System, we first need to know the current resonant populations and orbital distributions.  Identifying members of particular resonances is straightforward \citep[e.g.,][]{Chiang2003,Elliot2005,Lykawka2007,Gladman2008,Volk2011}, but using the observed set of resonant TNOs to infer the intrinsic number and distribution of resonant objects is difficult due to complicated observational biases induced by the resonant orbital dynamics \citep{Kavelaars2009,Gladman2012}.  Here, we present the first set of 31 secure and 8 insecure resonant TNOs detected by the Outer Solar System Origins Survey (OSSOS), which was designed to produce detections with well-characterized biases \citep{Bannister2015}.  

OSSOS is a Large Program on the Canada-France Hawaii Telescope surveying eight $\sim21$ deg$^2$ fields, some near the invariable plane and some at moderate latitudes from the invariable plane, for TNOs down to a limiting magnitude of $\sim24.5$ in r-band.  Observations began in spring 2013 and will continue through early 2017 (see \citealt{Bannister2015} for a full description of OSSOS).  Two of the primary science goals for OSSOS are measuring the relative populations of Neptune's mean motion resonances and modeling the detailed orbital distributions inside the resonances.  The most current observational constraints on both the distributions and number of TNOs in Neptune's most prominent resonances come from the results of the Canada France Ecliptic Plane Survey \citep[CFEPS;][hereafter referred to as G12]{Gladman2012}. Population estimates for some of Neptune's resonances have also been modeled based on the Deep Ecliptic Survey \citep[DES;][]{Adams2014}.   OSSOS will offer an improvement on these previous constraints because it is optimized for resonant detections (especially for the 3:2 resonance) and includes off-invariable plane blocks to better probe inclination distributions. 

\begin{figure}[h]
   \centering
   \includegraphics[width=3.2in]{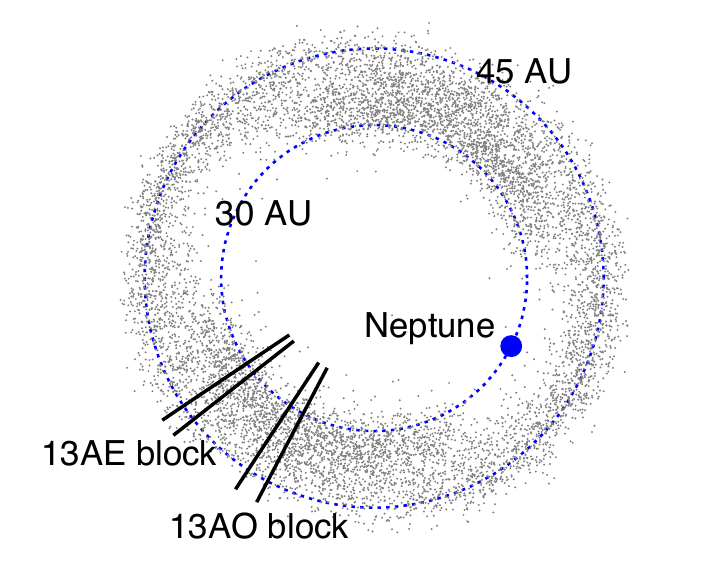}
\caption{Top-down view of the Solar System showing the locations of the 13AO and 13AE OSSOS blocks relative to the 3:2 resonance model from G12.  The 13AO block is off the invariable plane near the trailing ortho-Neptune point and the 13AE block straddles the ecliptic and invariable planes $20^{\circ}$ farther from Neptune.}
\label{f:OE_blocks}
\end{figure}

Here we report on the characterized resonant object detections from the first two of the eight OSSOS observational blocks: 13AO\footnote{The 13A designation indicates that the discovery images for these blocks were observed at opposition in CFHT's 2013 A semester.}, an off-invariable plane block with a characterization limit of $m_r=24.39$, and 13AE, a block overlapping the ecliptic and invariable planes with a characterization limit of $m_r=24.04$. The characterization limit is the faintest magnitude for which the detection efficiency of the survey is well-measured and for which all objects are tracked (see \citealt{Bannister2015} for more details).  Figure~\ref{f:OE_blocks} shows the location of the 13AO and 13AE blocks relative to a model (G12) of Neptune's 3:2 mean motion resonance.  The 13AO block is centered about $7^{\circ}$ above the ecliptic plane at the trailing ortho-Neptune point ($90^{\circ}$ in longitude behind Neptune), and the 13AE block is at $0-3^{\circ}$ ecliptic latitude $\sim20^{\circ}$ farther from Neptune.  The full description of these blocks and the OSSOS observational methods are detailed in \citet{Bannister2015}.  The 13AO block yielded 18 securely resonant TNOs out of 36 characterized detections, and the 13AE block yielded 13 securely resonant TNOs out of 50 characterized detections; securely resonant objects are ones where the orbit-fit uncertainties fall within the width of the resonance (see Section~\ref{s:detections} and Appendix~\ref{a:orbits} for the full list of resonant OSSOS detections and discussion of the classification procedure).  The larger yield of resonant objects in the 13AO block reflects both its favorable placement off the invariable plane near the center of the 3:2 resonance (see Section \ref{s:32}) as well as its slightly fainter characterization limit.

Our detections include secure  3:2, 5:2, 2:1, 7:3, and 7:4 resonant objects as well as insecure 5:3, 8:5, 18:11, 16:9, 15:8, 13:5, and 11:4 detections.  The 3:2, 5:2, and 2:1 resonances contain sufficiently many detections to model their populations.  In this paper, we use these detections and the survey's known biases to place constraints on the number and distribution of objects in these resonances. We then discus how our constraints compare to the current theoretical understanding of the origins and dynamics of these TNOs. 

Current models propose three possible pathways by which TNOs may be captured into resonance.  First, they may have been captured by Neptune as it smoothly migrated outward on a roughly circular orbit \citep{Malhotra1993,Malhotra1995,Hahn2005}, driven by interactions between Neptune and a primordial planetesimal disk \citep{Fernandez1984}.  A disk with the majority of its mass in planetesimals $<$100km in radius can produce migration smooth enough for resonance capture  \citep{Murray-Clay2006}, although formation of large planetesimals by the streaming instability \citep{Youdin2005,Johansen2007}, if efficient, could render planetesimal-driven migration too stochastic.  Capture by a smoothly migrating Neptune produces some objects that are deeply embedded  in the resonance, having resonant angles (see Section \ref{s:methods}) that librate with low amplitude \citep[e.g.,][]{Chiang2002}. Given capture by smooth migration, the distribution of libration centers (see Section \ref{s:21}) among 2:1 resonant objects serves as a speedometer, measuring the timescale of Neptune's primordial orbital evolution \citep{Chiang2002,Murray-Clay2005}.  Smooth migration models predict that the 5:2 resonance captures fewer objects than the 3:2 and 2:1 \citep{Chiang2003,Hahn2005} and have difficulty producing the large inclinations observed in the Kuiper belt, though \citet{Nesvorny2015} recently suggest that transient resonant sticking and loss during slow migration may resolve the latter difficulty.

Second, resonant objects could be the most stable remnants of a dynamically excited population that filled phase space in the outer Solar System \citep{Levison2008,Morbidelli2008} as a result of early dynamical instability among the giant planets \citep[e.g.,][]{Thommes1999,Tsiganis2005}.  The phase space volume of each resonance in which objects can have small libration amplitudes is limited, so this type of model preferentially produces larger-amplitude librators.  Because Neptune spends time with high eccentricity, such a scenario must be tuned to avoid disruption of the observed dynamically unexcited ``cold classical" TNOs \citep{Batygin2011,Wolff2012,Dawson2012}.  Models of capture following dynamical instability may produce high inclination TNOs more effectively than standard smooth migration models, but they still under-predict observations \citep{Levison2008}. Like smooth migration, these models do not predict a large 5:2 population compared to the 3:2 and 2:1 populations \citep[G12]{Levison2008}. 

Third, resonant objects need not be primordial.  Objects currently scattering off of Neptune can be captured into resonance temporarily \citep[e.g.,][]{Lykawka2007,Pike2015}.  These marginally stable objects tend to have large libration amplitudes and may be a productive source of objects in distant resonances such as the 5:2.  

Inspired by the differences between these three emplacement mechanisms, we focus our dynamical modeling on the libration amplitude distribution in the 3:2 resonance, the distribution of libration centers in the 2:1 resonance, and the relative abundance of objects in the 5:2 compared to the 3:2 and 2:1.  Finally, we emphasize that comparison of dynamical models of resonance capture with the current resonant populations must take into account the evolution of resonant orbits over the age of the Solar System.  Numerous theoretical studies of the current dynamics and stability of Neptune's resonances \citep[e.g.,][]{Gallardo1998,Yu1999,Nesvorny2000,Nesvorny2001,Tiscareno2009} provide insight into this evolution.  We use these studies to inform our constructed orbital models.

We provide a brief description of Neptune's resonances as well as our methods for modeling the resonant populations in Section \ref{s:methods}.  A table of the OSSOS resonant TNO detections from the 13AE and 13AO blocks and a description of how we determine resonance membership is provided in Section \ref{s:detections}.  We model the distribution and total number of 3:2 objects based on the OSSOS detections in Section \ref{s:32}, demonstrating that this population contains members with lower libration amplitudes than previously seen.   In Section \ref{s:52}, we model the 5:2 resonance and confirm the unexpectedly large population of 5:2 resonant objects reported in G12.  Section \ref{s:21} presents a model of the 2:1 resonance and provides an improved constraint on the relative number of symmetric and asymmetric librators.  We summarize our population estimates and compare with previous results in Section \ref{s:pop}, and we comment on the populations of resonances for which we have insecure detections in Section \ref{s:other}.  Section \ref{s:sum} summarizes our key results.

\section{Background and methods}\label{s:methods}

After briefly introducing mean motion resonances (Section \ref{s:mmr}), we summarize the challenges presented by detection bias for measuring resonant populations (Section \ref{s:bias}). To circumvent these biases, we employ the OSSOS survey simulator to test models of the resonant populations.  The survey simulator can only test models---it cannot produce them---and we describe our choice of models in  Section \ref{ss:ssmod}.

\subsection{Neptune's mean motion resonances} \label{s:mmr}

Neptune's mean motion resonances (which we call $p:q$ resonances, with $p>q>0$ for external resonances) have resonant angles, $\phi$, given by
\begin{equation}\label{eq:full_phi} 
\phi = p\lambda_{tno} - q\lambda_{N} - r_{tno}\varpi_{tno}-r_{N}\varpi_{N}-s_{tno}\Omega_{tno}-s_{N}\Omega_{N}
\end{equation} 
where $\lambda,\varpi$, and $\Omega$ are the mean longitude, longitude of perihelion, and longitude of ascending node (the subscripts $tno$ and $N$ refer to the elements of a TNO and Neptune), and $p,q,r_{tno},r_N,s_{tno},s_N$ are integers with the constraint that $p - q - r_{tno}-r_{N}-s_{tno}-s_{N}=0$.  Objects in a mean motion resonance have values of $\phi$ that librate around a central value with an amplitude defined as $A_\phi = (\phi_{max} -\phi_{min})/2$.  For small eccentricity ($e$) and inclination ($i$), the strength of the resonant terms in the disturbing function are proportional to 
$e_{tno}^{|r_{tno}|}e_N^{|r_{N}|}(\sin i_{tno})^{|s_{tno}|}(\sin i_{N})^{|s_{N}|}$ 
\citep{MurrayDermott1999}, and resonances with small $|p-q|$ are generally stronger than those with larger $|p-q|$.  TNOs typically have eccentricities and inclinations much larger than Neptune's, so we will ignore resonant angles involving $\varpi_{N}$ and $\Omega_{N}$.  Likewise the resonant angles involving the inclination of the TNO are typically less important than those involving the eccentricity because inclination resonances are at least second order in $\sin i_{tno}$.  Throughout the rest of this work, we will generally consider this simplified resonance angle:
\begin{equation}\label{eq:phi}
\phi = p \lambda_{tno} - q \lambda_{N} - (p-q)\varpi_{tno}
\end{equation}
with a few exceptions noted in Table~\ref{t:detections} and Section~\ref{s:other}.  In most cases, such as in the 3:2 and 5:2 resonances, this resonant angle librates around $\phi = 180^{\circ}$. The topology of n:1 exterior resonances allows for resonant orbits with more than one center of libration; the 2:1 resonance has two so-called asymmetric libration centers near $\phi\sim60-100^{\circ}$ and $\phi\sim260-300^{\circ}$ (the exact centers are eccentricity dependent) in addition to the symmetric libration center at $\phi = 180^{\circ}$.  The libration of $\phi$ around specific values means that objects in resonance will come to perihelion at specific offsets from Neptune's current mean longitude.  When a TNO is at perihelion, its mean anomaly ($M$) is 0, so $\lambda_{tno} = M + \varpi = \varpi$. Substituting this into equation~\ref{eq:phi} shows that at perihelion:
\begin{equation}\label{eq:peri}
\varpi - \lambda_{N} =  \lambda_{tno} -  \lambda_{N} = \frac{\phi}{q}.
\end{equation}

Some resonances contain a subcomponent of objects also in the Kozai resonance; these objects exhibit libration of the argument of perihelion, $\omega = \varpi - \Omega$, in addition to libration of the resonant angle $\phi$.  This libration causes coupled variations in $e$ and $i$ such that the quantity $\sqrt{1-e^2}\cos i$ is preserved.  Outside of mean motion resonances, libration of $\omega$ only occurs at very large inclinations in the trans-Neptunian region \citep{Thomas1996}, but inside mean motion resonances Kozai libration can occur at much smaller inclinations.  In the 3:2 resonance, Kozai libration can occur even at very low inclinations \citep{Morbidelli1995} and a significant number of observed 3:2 objects are known to be in the Kozai resonance, including Pluto.  Kozai resonance has also been observed for members of the 7:4, 5:3, and 2:1 resonances \citep{Lykawka2007}.  In the 3:2 resonance, the libration of $\omega$ occurs around values of $90^{\circ}$ and $270^{\circ}$ with typical amplitudes of $10-70^{\circ}$ and typical libration periods of several Myr.

\subsection{Detection biases for resonant objects} \label{s:bias}

In order to be detected by OSSOS, a TNO must be in the survey's field of view, brighter than the limiting magnitude of the field, and moving at a rate of motion detectable by the survey's moving object detection pipeline (see \citealt{Bannister2015} for more details); because the OSSOS observing strategy is optimized to detect the motion of objects at distances between $\sim9-300$ AU, the first two criteria are the primary source of detection biases for the resonant objects.  The intrinsic brightness distribution of TNOs with absolute magnitudes brighter than $H_r\sim8$ is generally well-modeled as an exponential in $H$ (discussed in Section~\ref{ss:ssmod}), meaning there are increasing numbers of objects at increasing $H$ (decreasing brightness).  For a population of TNOs on eccentric orbits, this means that most detections will be made for faint, large-$H$ TNOs near their perihelion.  Consequently, populations containing preferentially fainter objects must have preferentially higher eccentricities to produce the same number of detections.  Furthermore, given that resonant TNOs come to perihelion at preferred longitudes relative to Neptune (equation~\ref{eq:peri}), this means that the placement of the field in longitude relative to Neptune produces biases toward and against certain resonances. Objects in n:2 resonances librating about $\phi=180^{\circ}$ will preferentially come to perihelion at the ortho-Neptune points ($\pm90^{\circ}$ away from Neptune); asymmetric n:1 librators will come to perihelion at various longitudes ahead or behind Neptune, depending on the value of the libration center (see Figure~1 in G12 for an illustration of perihelion locations for various resonances). The OSSOS 13AO and 13AE blocks are $\sim90^{\circ}$ and $\sim110^{\circ}$ behind Neptune, which favors the detection of n:2 objects as well as asymmetric librators in the 2:1 resonance's trailing libration center.  These biases for the 3:2, 5:2, and 2:1 resonances will be discussed in later sections.  

Similarly, latitude placement of the observing blocks relative to the ecliptic plane produces biases in inclination for TNOs.  The 13AE block ($0-3^{\circ}$ ecliptic latitude) favors the detection of low-$i$ TNOs because these TNOs spend most of their time near the ecliptic plane, while in the 13AO block it is not possible to detect objects with inclinations smaller than the field's ecliptic latitude of $6-9^{\circ}$.  For resonances such as the 3:2, the Kozai subcomponent of the resonance introduces an additional observational bias; the libration of $\omega$ means that Kozai resonant objects come to perihelion at preferred ecliptic latitudes in addition to preferred longitudes with respect the Neptune (equation~\ref{eq:peri}).  The biases induced by the Kozai resonance for the 3:2 population are discussed in detail by \citet{Lawler2013}. To account for these observational biases in our modeling, we use the OSSOS survey simulator.

\subsection{Modeling Neptune's resonances using a survey simulator}\label{ss:ssmod}

We use the OSSOS detections of resonant objects combined with the OSSOS survey simulator\footnote{\url{https://github.com/OSSOS/SurveySimulator}} to construct and test models of Neptune's resonant populations.  The survey simulator is described in \citet{Bannister2015}.  Its premise is as follows: given a procedure (i.e model) for generating the position and brightness of resonant objects on the sky, the simulator repeatedly generates objects and then checks whether they would have been detected by the survey.  The simulator stops when the desired number of simulated detections is achieved.  When the model agrees with observations, the sets of real and simulated detected objects should have similar absolute magnitudes and orbital properties.  The intrinsic number of objects in a resonance (i.e. a population estimate for the input model) corresponds to the number of detected and undetected objects the survey simulator had to generate (down to a specified absolute magnitude $H$) in order to match the real number of detections.  We run the survey simulator many times for each model with different random number generator seeds; this allows us to build a distribution of population estimates and a large sample of simulated detections.  We then run statistical tests to determine whether the model provides simulated detections that are a good match to the real detections; these tests are discussed later in this section as well as in Appendix~\ref{a:stats}.  

A resonant object's orbit is uniquely determined by its semi-major axis, $a$, eccentricity, $e$, inclination, $i$, mean anomaly, $M = \lambda - \varpi$, longitude of ascending node, $\Omega$, resonance angle $\phi$, and epoch, $t$, for the given value of $M$.  Following G12, we construct a set of models for each resonant population by parameterizing the intrinsic distributions of $a$, $e$, $i$, $\phi$, and absolute magnitude $H$. For each simulated object, the simulator draws $a$, $e$, $i$, $\phi$, and $H$ from these models and then constructs the remaining orbital elements based on constraints from the resonant condition (equation~\ref{eq:phi}).  We choose a uniformly-distributed random value for $M$ to reflect that the object's specific position within its orbit is random in time, and we draw $a$ randomly from a uniform distribution spanning the approximate resonance width.  Appendix~\ref{a:resmodeling} provides the values used for the resonance widths, though we note that our results are not affected by this complication; because the resonance widths are small, choosing a fixed $a$ for each resonance would produce equivalent results.  For objects not experiencing Kozai oscillations, the orientation of the orbit's plane relative to the ecliptic plane is not coupled to the resonance, so we also choose $\Omega$ randomly from a uniform distribution.  For the 3:2 population, we include an additional parameter for the fraction of the population in the Kozai resonance.  Our procedure for selecting the orbital elements of these objects is described in Section \ref{s:32} and Appendix \ref{as:32}.

In Sections~\ref{s:32} through~\ref{s:21} and Appendix~\ref{a:resmodeling} we outline the exact models used, but the general form of the parameterized models in $H$, $e$, and $i$ is the same for each resonance.  We represent the cumulative luminosity distribution as an exponential in $H$ with logarithmic slope $\alpha$:
\begin{equation}\label{eq:spl}
N(<H) = 10^{\alpha (H-H_0)},
\end{equation}
where $N(<H)$ is the number of objects having magnitudes between a reference $H_0$ and $H$.  This form models the absolute magnitude distribution well for $H_r \lesssim 8$ \citep[e.g.,][G12]{Fraser2009,Fuentes2009,Shankman2013,Fraser2014}, but is not expected to work well for intrinsically fainter objects (see Section \ref{ss:32Hd}).

We model the differential eccentricity distribution as a Gaussian centered on $e_c$ with a width $\sigma_e$:
\begin{equation} \label{eq:e}
\frac{dN(e)}{de} \propto \exp{\left(-\frac{(e-e_c)^2}{2\sigma_e\,^2}\right)},
\end{equation}
where $dN(e)$ is the number of objects with eccentricities between $e$ and $e + de$.  This is a convenient form that acceptably describes populations with a typical eccentricity and a roughly symmetrical eccentricity dispersion.
Following \citet{Brown2001}, we model the differential inclination distribution as a Gaussian with width $\sigma_i$ multiplied by $\sin(i)$:
\begin{equation} \label{eq:i}
\frac{dN(i)}{di} \propto \sin(i) \,\exp{\left(-\frac{i^2}{2\sigma_i\,^2}\right)},
\end{equation}
where $dN(i)$ is the number of objects with inclinations between $i$ and $i + di$.

The $\phi$ distribution and treatment of the Kozai resonance are specific to each resonance.  However for the 3:2 and 5:2 resonances,  which have only one libration center $\phi = 180^{\circ}$, the $\phi$ distribution may be uniquely specified by a distribution of libration amplitudes, $A_\phi$, about that center.  We approximate the time evolution of $\phi$ for an individual object as the oscillation of a simple harmonic oscillator with amplitude $A_\phi$ \citep{MurrayDermott1999}.  The instantaneous value of $\phi$ for a simulated object is then  
\begin{equation} \label{eq:phitime}
\phi  = \phi_{center} + A_{\phi} \sin(2\pi t),
\end{equation}
where $t$ is a random number distributed uniformly between 0 and 1.  Small-amplitude libration is well-approximated by a simple harmonic oscillator, while for large $A_\phi$ the angular evolution near the extrema of libration (where $\dot \phi$ changes sign) slows less in full numerical simulations than equation~\ref{eq:phitime} implies.  This means that compared to full numerical simulations, equation~\ref{eq:phitime} slightly underestimates the likelihood that objects will be observed $90^\circ$ from Neptune (perihelion for $\phi = 180^\circ$) and slightly overestimates the likelihood of finding objects at angles corresponding to the extrema of libration.  However, in Appendix \ref{as:32} we demonstrate that for all plutinos observed by OSSOS, full simulations of the resonant angle evolution do not deviate from equation~\ref{eq:phitime} enough to meaningfully affect our results.

For resonances with a single libration center, we follow G12 and model the distribution of libration amplitudes as a triangle that starts at $A_{\phi, min}$, rises linearly to a central value $A_{\phi, c}$ and then linearly falls to zero at the upper stability boundary for $A_{\phi, max}$ ($\sim150^{\circ}$ in the case of the 3:2, \citealt{Tiscareno2009}).  This triangle need not be symmetric.    A triangular $A_{\phi}$ distribution is not an arbitrary choice; theoretical studies of resonant phase space and of the dynamical capture and the evolution of plutinos often result in $A_{\phi}$ distributions that are roughly triangular in shape \citep[e.g.,][]{Nesvorny2000,Chiang2002,Lykawka2007}.  This outcome may be understood qualitatively as the result of shrinking phase space volumes at small libration amplitudes and increased dynamical instability at large amplitudes.  For example, plutinos with $A_{\phi}\gtrsim120^{\circ}$ are not stable on Gyr timescales \citep{Nesvorny2000,Tiscareno2009}.

When comparing real and simulated detections, we consider the following observables for each object: absolute magnitude $H$, eccentricity $e$, inclination $i$, heliocentric distance at detection $d$, and libration amplitude $A_{\phi}$ (see Section~\ref{s:detections} and Appendix~\ref{a:orbits} for discussion of how $A_{\phi}$ is determined for the observed objects).  Because we are modeling each resonance separately, we do not consider the semimajor axis distribution within the resonance; the small variations in $a$ for each object compared to the exact resonant value do not affect observability, so the $a$ distribution is not a useful model test.  For plutinos, we also compare the observed and simulated fraction of objects in the Kozai resonance (see Section~\ref{ss:32kz}).  For resonances such as the 2:1 with symmetric and asymmetric libration centers, we compare the observed and simulated fractions of objects in each libration island (see Section~\ref{s:21}).  We do not compare the $\Omega$, $M$, or $\varpi$ distributions of the simulated and real detections because these angles are related by $\phi$.  We also do not compare the $\omega$ distributions because this angle is evenly distributed except in the case of Kozai resonance.

For each model test, we have two goals: (1) to determine the range of model parameters that provide acceptable matches with the data and (2) to determine the model parameters that best fit the data.  We note that our statistical approach is limited by computational feasibility. Ideally we would like to perform a maximum likelihood calculation, but the nature of the observational biases means we cannot analytically calculate the detection probabilities; they must instead be numerically determined by running the survey simulator.  Given the wide range of possible models for the populations we are investigating, using the survey simulator to perform a maximum likelihood calculation is not currently feasible (see Appendix~\ref{a:stats} for a detailed explanation). 

The observational biases affecting the $i$ and $A_{\phi}$ distributions are relatively independent of each other and of the chosen $H$ and $e$ distributions; to reduce the complexity of model testing, we consider each of these observables separately as a one-dimensional distribution.  Following \citet{Petit2011} and G12, we use the Anderson-Darling (AD) test to identify a range of acceptable model parameters for these two distributions (goal 1 above).  The test statistic---described in Appendix~\ref{a:stats}---is a weighted measure of the difference between two cumulative distributions.  For each set of model parameters, we determine whether the set of $i$ and $A_{\phi}$ values for the real detections could be drawn from the simulated detections as follows:  we generate a large number of synthetic detections for each model and calculate the AD statistic for the real detections compared to the model distribution.  We then determine the significance of that value of the AD statistic for the $N$ real objects by randomly drawing subsamples of $N$ synthetic detections from the model detections and calculating the AD statistic for these subsamples (i.e. bootstrapping).  We reject a model if the AD statistic for the real detections is larger than the AD statistic for $95\%$ or more of the model subsamples compared to the model itself.  This procedure yields our $95\%$ confidence limits on the acceptable parameters ($\sigma_i$ and $A_{\phi, c}$) for our orbital model.  We note that this bootstrapping is required to produce confidence limits from the AD statistic because our distributions are not gaussian.  The AD test is a model rejection test, so to get a most probable values of $\sigma_i$ and $A_{\phi, c}$ (goal 2), we must employ a different procedure.  We numerically generate one-dimensional probability distributions in $i$ and  $A_{\phi}$ for each allowed value of the parameters $\sigma_i$ and $A_{\phi, c}$, calculate the probability of detecting the observed objects for each parameter value, and select the values of $\sigma_i$ and $A_{\phi, c}$ that maximize this probability.  See Appendix~\ref{a:stats} for more details on how this calculation is done.  We note that a bootstrapping procedure to determine the significance of the calculated probabilities of $\sigma_i$ and $A_{\phi, c}$ yield 95\% confidence limits on those values that are very similar to the 95\% confidence limits based on the AD statistic.

The observational biases that affect the $H$, $e$, and $d$ distributions are coupled such that the best parameters for the $H$ and $e$ distributions cannot be determined independently of each other (see discussions in \citealt{Kavelaars2009} and G12).  In these cases we calculate the one-dimensional AD statistic for the observed $H$, $e$, and $d$ distributions compared to the model's synthetic detections. Following \citet{Parker2015} and \citet{Alexandersen2015}, instead of calculating the significance of each of these statistics individually, we calculate the significance of the sum of the observed distribution's $H$, $e$, and $d$ AD statistics relative to the same sum for the model compared to itself.  We reject combinations of $\alpha$, $\sigma_e$, and $e_c$ for which the summed AD statistic is larger than $95\%$ of the summed statistics for the subsets of synthetic detections.  To determine our preferred values of $\alpha$, $\sigma_e$, and $e_c$ within those $95\%$ confidence limits, we use the sum of their one-dimensional $\chi$-square values and select the $\alpha$, $\sigma_e$, and $e_c$ that minimizes this sum.  These values will not necessarily be the true, most probable values of $\alpha$, $\sigma_e$, and $e_c$ for our parameterized models because the one-dimensional $\chi$-square values do not account for how well the data fits the model in three-dimensional $H$, $e$, and $d$ space.  Correctly determining the most probable values of $\alpha$, $\sigma_e$, and $e_c$ is computationally too expensive for the wide range of parameter space we must explore (see discussion in Appendix~\ref{a:stats}).

\section{OSSOS Resonant Detections}\label{s:detections}

We use the classification scheme outlined in \citet{Gladman2008} to determine which OSSOS detections are resonant: a best-fit orbit for each OSSOS detection is computed using the \citet{Bernstein2000} algorithm and then a search around the best-fit orbit is done to find the maximum and minimum acceptable semimajor axis orbits.  Following \citet{Gladman2008}, an orbit is deemed an acceptable fit to the observations if it meets two conditions: (1) the worst residual when comparing the observed astrometric position of the objects to the positions predicted by the orbit are not more than 1.5 times the worst residual for the best-fit orbit and (2) the rms residual is not more than 1.5 times the best-fit orbit's rms residual.  The best-fit, minimum-$a$, and maximum-$a$ orbits are integrated forward in time to look for resonant behavior (defined as libration of a resonance angle described by equation~\ref{eq:full_phi}) on $10^7$ year timescales; we check all potential resonances with $|p-q| \le 30$ within 2\% of the best-fit orbit's semimajor axis.  The resonant objects usually require more precise orbit fits than non-resonant objects in order to achieve secure classifications (meaning all three orbits are resonant) because uncertainty in an object's semimajor axis leads to uncertainty in the libration amplitude.  This classification procedure yielded 21 secure 3:2 objects, 4 secure 2:1 objects, 4 secure 5:2 objects, 1 secure 7:4 object, and 1 secure 7:3 object.  We also have 2 insecure 5:3 objects, and 1 insecure detection in each of the 11:4, 8:5, 18:11, 16:9, 15:8, and 13:5 resonances.  These objects are listed in Table~\ref{t:detections} along with their best-fit orbital parameters with uncertainties. The listed uncertainty in $a$ is the $1-\sigma$ uncertainty calculated from the \citet{Bernstein2000} orbit-fit covariance matrix; the uncertainties in $e$ and $i$ are all small and rather than list them, these parameters have been reported to the appropriate number of significant figures.  The uncertainty in the libration amplitude is obtained by integrating 250 clones of each object's best-fit orbit (obtained from the covariance matrix), measuring their $A_{\phi}$ distribution, and calculating the $1-\sigma$ (68\%) confidence range; see Appendix~\ref{a:orbits} for a full discussion of the classification scheme and determination of the $A_{\phi}$ distributions.  Many of the best-fit orbits' $A_{\phi}$ distributions are asymmetric around the best-fit orbit's $A_{\phi}$, and in these cases the `$1-\sigma$' uncertainties listed in Table~\ref{t:detections} actually represent hard upper or lower limits to the value of $A_{\phi}$ (see Figure~\ref{af:aphi-dist} in Appendix~\ref{a:orbits}); these instances are marked in the table by asterisks.  We note that in many cases the orbit-fit uncertainties, and especially the libration amplitude uncertainties, are quite small even though the total arc length on the observations is only $\sim17$ months; this is due to the optimized observing schedule and accurate astrometry \citep{Bannister2015}.  The libration amplitude uncertainties listed in Table~\ref{t:detections} are comparable to or smaller than those determined for other TNOs with significantly longer observational arcs (see, for example, \citealt{Lykawka2007}).

\begin{deluxetable}{l l l l l r l l l l l}
\rotate
\tabletypesize{\footnotesize}
\tablecolumns{11}
\tablewidth{0pt}
\tablecaption{}
\tablehead{\multicolumn{2}{c}{Designations} & Res & \colhead{a} & \colhead{e} & \colhead{i} & \colhead{d} & \colhead{$A_{\phi}$}& \colhead{Mag}& \colhead{H} & \colhead {Comment} \\ 
OSSOS & MPC & \colhead{ } & \colhead{(AU)} & \colhead{}& \colhead{($^{\circ}$)} & \colhead{(AU)} & \colhead{($^{\circ}$)} & \colhead{(r)} & \colhead{(r)} & \colhead{ } }
\startdata
o3e02	&	2013 GH137	& 3:2		& $39.441\pm0.013$		&	0.2282	& 13.468	& 31.1	&	$\phm{1}67^{+3}_{-2}$	& 23.4	& 8.32	& \\[2pt]
o3e03 	&	2013 GE137	& 3:2		& $39.332\pm0.028$		&	0.257	& 3.866	& 31.1	&	$\phm{1}81^{+7}_{-5}$	& 23.8	& 8.70	& \\[2pt]
o3e04	&	2013 GJ137	& 3:2		& $39.466\pm0.033$		&	0.265	& 16.873	& 32.1	&	$\phm{1}61^{+10}_{-8*}$	& 23.5	& 8.25	& Kozai: $270\pm40^{\circ}$\\[2pt]
o3e06	&	2013 GL137	& 3:2		& $39.249\pm0.027$		&	0.199	& 10.440	& 34.4	&	$101^{+12}_{-10}$		& 23.8	& 8.52	& \\[2pt]
o3e07	&	2013 GG137	& 3:2		& $39.340\pm0.019$		&	0.136	& 2.931	& 35.2	&	$\phm{1}65^{+6}_{-6}$	& 24.2	& 8.52	& \\[2pt]
o3e08	&	2013 GD137	& 3:2		& $39.371\pm0.005$		&	0.1035	& 6.943	& 35.4	&	$\phm{1}69^{+2}_{-1}$	& 23.7	& 8.45	& \\[2pt]
o3e12	&	2013 GF137	& 3:2		& $39.557\pm0.011$		&	0.1567	& 14.680	& 37.3	&	$\phm{1}98^{+3}_{-6}$	& 23.7	& 8.11	& \\[2pt]
o3e41	&	2013 GK137	& 3:2		& $39.168\pm0.015$		&	0.178	& 9.879	& 45.6	&	$140^{+15}_{-10}$		& 23.8	& 7.28	& \\[2pt]
o3o02	&	2013 JC65	& 3:2		& $39.363\pm0.018$		&	0.2939	& 16.409	& 28.2	&	$\phm{1}38^{+9}_{-7*}$	& 23.8	& 9.11	& Kozai: $\phm{2}90\pm50^{\circ}$\\[2pt]
o3o03	&	2013 JH65	&3:2		& $39.340\pm0.030$		&	0.286	& 7.533	& 29.5	&	$\phm{1}92^{+25}_{-14}$	& 24.0	& 9.22	& \\[2pt]
o3o04	&	2013 JG65	&3:2		& $39.362\pm0.015$		&	0.2492	& 15.934	& 30.0	&	$\phm{1}34^{+8}_{-6}$	& 23.7	& 8.79	& Kozai: $\phm{2}90\pm70^{\circ}$\\[2pt]
o3o05	&	2013 JK65	&3:2		& $39.411\pm0.017$		&	0.2564	& 20.045	& 30.1	&	$\phm{1}10^{+8}_{-4*}$	& 24.3	& 9.44	& \\[2pt]
o3o06	&	2013 JZ64	&3:2		& $39.416\pm0.015$		&	0.2325	& 10.128	& 30.4	&	$\phm{1}18^{+2}_{-2*}$		& 23.8	& 8.82	& Kozai: $\phm{2}90\pm60^{\circ}$\\[2pt]
o3o08	&	2013 JE65	&3:2		& $39.350\pm0.041$		&	0.278	& 8.048	& 31.7	&	$\phm{1}42^{+10}_{-10*}$		& 24.3	& 9.15	& Kozai: $\phm{2}90\pm70^{\circ}$\\[2pt]
o3o09	&	2013 JB65	&3:2		& $39.403\pm0.005$		&	0.1889	& 24.898	& 32.0	&	$\phm{1}26^{+2}_{-1*}$	& 23.4	& 8.13	& \\[2pt]
o3o10	&	2013 JF65	&3:2		& $39.393\pm0.007$		&	0.1764	& 8.315	& 32.5	&	$\phm{1}16\pm3$		& 24.2	& 8.94	& \\[2pt]
o3o12	&	2013 JA65	&	3:2	& $39.520\pm0.009$		&	0.1488	& 10.223	& 33.8	&	$\phm{1}75^{+3}_{-8}$	& 24.2	& 8.78	& \\[2pt]
o3o13	&	2013 JL65		&	3:2	& $39.282\pm0.053$		&	0.230	& 7.251	& 34.7	&	$\phm{1}72^{+24}_{-32}$	& 24.4	& 8.79	& \\[2pt]
o3o15	&	2013 JD65	&	3:2	& $39.371\pm0.006$		&	0.0937	& 13.015	& 35.7	&	$\phm{1}50^{+2}_{-3}$	& 23.7	& 7.90	& \\[2pt]
o3o20PD	& 	2007 JF43	&	3:2	& $39.381\pm0.071$		&	0.186	& 15.080	& 38.3	&	$\phm{1}48^{+19}_{-5*}$	& 21.2	& 5.27	& \\[2pt]
o3o27	&	2013 JJ65		&	3:2	& $39.391\pm0.063$		&	0.256	& 19.814	& 41.0	&	$\phm{1}28^{+21}_{-1}$	& 23.5	& 7.22	& \\[2pt]
\hline \\[-7pt]
o3e05	& 2013 GW136	&	2:1	& $47.741\pm0.015$		&	0.3440	& 6.660	& 33.0	&	$\phm{1}41\pm2$		& 22.7	& 7.42	& asym. $\phi_c=278^{\circ}$\\[2pt]
o3e55	& 2013 GX136	&	2:1	& $48.011\pm0.013$		&	0.2519	& 1.100	& 37.0	&	$157\pm1$			& 23.9	& 7.67	& \\[2pt]
o3o18	& 2013 JE64	&	2:1	& $47.762\pm0.059$		&	0.284	& 8.335	& 36.1	&	$\phm{1}21^{+8}_{-1*}$	& 23.6	& 7.94	& asym. $\phi_c=285^{\circ}$\\[2pt]
o3o33	& 2013 JJ64	&	2:1	& $47.766\pm0.033$		&	0.082	& 7.650	& 41.0	&	$133^{+7}_{-3}$		& 24.0	& 7.27	& \\[2pt]
\hline \\[-7pt]
o3e09	& 2013 GY136	&	5:2	& $55.549\pm0.031$		&	0.4143	& 10.877	& 35.8	&	$\phm{1}88\pm10$		& 23.1	& 7.32	& \\[2pt]
o3e48	& 2013 GS136	&	5:2	& $55.629\pm0.034$		&	0.3855	& 6.978	& 35.5	&	$122\pm10$			& 23.9	& 8.50	& \\[2pt]
o3o07	& 2013 JF64	&	5:2	& $55.423\pm0.014$		&	0.4497	& 8.785	& 30.5	&	$\phm{1}62\pm9$		& 24.1	& 8.97	& \\[2pt]
o3o11	& 2013 JK64	&	5:2	& $55.238\pm0.037$		&	0.4081	& 11.078	& 33.1	&	$\phm{1}81\pm16$		& 23.0	& 7.69	& \\[2pt]
\hline \\[-7pt]
o3e19	& 2013 GR136	&	7:4	& $43.649\pm0.007$		& 0.0767 		& 1.645	& 41.0	&	$\phm{1}88^{+7}_{-11}$	& 23.8	& 7.20	& \\[2pt]
\hline \\[-7pt]
o3o19	& 2013 JN64	&    7:3	& $53.032\pm0.039$		& 0.287 		& 7.740	& 38.1	&	$128^{+28}_{-16}$		& 24.1	& 7.96	& \\[2pt]
\hline \\[-7pt]
o3e17	& 2013 GV136	& 8:5(I)	& $41.098\pm0.007$		& 0.035		& 7.452	& 40.6	&						& 24.3	& 7.85	& insecure\\&&&&&&&&&& $90\%$ of clones res for $>5$ Myr\\[2pt]
o3o32	& 2013 JG64	& 18:11(IH) & $41.740\pm0.026$	& 0.111		& 18.208	& 45.7	&						& 24.1	& 7.48	& insecure, mixed argument\\&&&&&&&&&& best-fit + 20\% of clones res \\ &&&&&&&&&& $\phi = 18\lambda_{tno} - 11 \lambda_N - 5 \varpi_{tno} - 2 \Omega_{tno}$ \\[2pt]
o3e52	& 2013 GT136	& 5:3(I)	& $42.370\pm0.044$		& 0.1540		& 12.112	& 48.9	&						& 24.1	& 7.11	& insecure \\&&&&&&&&&& 40\% of clones res for $>5$ Myr\\[2pt]
o3o25	& 2013 JM64	&   5:3(IH)	& $43.352\pm0.009$		& 0.047		& 7.287	& 40.6	&						& 24.1	& 7.98	& insecure \\&&&&&&&&&& 75\% of clones res for $>5$ Myr \\[2pt]
o3e13	& 2013 GU136	& 16:9(IH)$^o$ & $44.14\pm0.02$		&0.169		& 8.318	& 37.9	&						& 23.6	& 7.86	& insecure\\&&&&&&&&&& 50\% of clones res for $>5$ Myr \\[2pt]
o3e49	& 2013 HR156	& 15:8(I)	& $45.73\pm0.01$		& 0.1890		& 20.412	& 38.1	&	$90^{+50}_{-10}$		& 23.6	& 7.72	& insecure, mixed argument\\&&&&&&&&&& 85\% of clones res \\ &&&&&&&&&& $\phi = 15\lambda_{tno} - 8 \lambda_N - 5 \varpi_{tno} - 2 \Omega_{tno}$ \\[2pt]
o3o29	& 2013 JL64	& 13:5(IH)$^o$&$56.8\pm0.1$		& 0.368		& 27.671	& 41.5	&						& 23.6	& 7.03	& insecure\\&&&&&&&&&& 35\% of clones res \\[2pt]
o3o34	& 2013 JH64	&  11:4(I)	& $59.08\pm0.25$		& 0.382 		& 13.731	& 50.8	&	$\sim60$				& 22.8	& 5.60	& insecure\\&&&&&&&&&&  50\% of clones res \\[2pt]
\enddata
\tablecomments{The uncertainties in $a$ are the 1-$\sigma$ uncertainties from the \citet{Bernstein2000} orbit fit; the listed values of $e$ and $i$ are printed with the appropriate number of significant figures. Note that the orbits above are barycentric osculating elements. `I' after the resonant classification indicates an insecure classification and `H' indicates that the human operator overrode the classification code's initial classification (in most instances this is due to messy libration behavior that was not correctly identified as libration by the automated code). A superscript `o' ($^o$) after the resonant classification indicates that some of the OSSOS astrometric measurements were discarded for determining the orbit fit and classification because of poor astrometric conditions.  The uncertainties in $A_{\phi}$ are obtained by generating 250 clones from the best-fit orbit's covariance matrix, integrating these clones for 10 Myr, and determining the 1-$\sigma$ (68\%) $A_{\phi}$ values from the clones' $A_{\phi}$ distribution.  An asterisk (*) following the uncertainty means that because of the asymmetric nature of the $A_{\phi}$ distribution, the stated uncertainty is a hard upper or lower limit for $A_{\phi}$.  For insecure resonant objects we list the percentage of clones that were resonant; a note of the percentage that are resonant for $>5$~ Myr indicates that the clones are only intermittently resonant.  If all of an object's 1$-\sigma$ clones are resonant we list the libration amplitude and uncertainties; in cases where the best-fit orbit results in well-behaved libration for 10 Myr, we list that single value of $A_{\phi}$.  We do not list libration amplitudes for objects whose clones intermittently librate; $A_{\phi}$ is not well-defined in these cases but is presumably large.  See Appendix~\ref{a:orbits} for a detailed discussion of the classification system and orbit determination. The OSSOS designations of the objects indicate which block they were discovered in with `o3o' objects being discovered in the 13AO block and `o3e' objects discovered in the 13AE block. }
\label{t:detections}
\end{deluxetable}
\clearpage 


\section{Plutinos: Population model and libration amplitudes}\label{s:32}

There are 21 characterized 3:2 objects from the OSSOS 13AO and 13AE blocks (listed in Table~\ref{t:detections}).  Our sample of 3:2 objects is sufficiently large to place some constraints on the libration amplitude distribution of Plutinos.  This distribution is of interest because it is likely to reflect plutino capture histories (see Section~\ref{s:intro}).  We begin by presenting maps illustrating the sensitivity of the OSSOS 13AE/O blocks as a function of phase space in the 3:2 resonance (Section \ref{ss:sensitivity}).  In the following sections, we use the survey simulator to constrain a parameterized model of the underlying 3:2 population.  The population's $i$ and $A_\phi$ distributions are modeled independently (Section \ref{ss:32iaphi}), while $H$ and $e$ must be constrained together (Section \ref{ss:32Hd}).  Section \ref{ss:32kz} presents constraints on the Kozai fraction, and we summarize and report a population estimate for the plutinos in Section \ref{ss:32sum}.

 \subsection{Sensitivity Maps}\label{ss:sensitivity}

\begin{figure}[htpb]
   \centering
   \includegraphics[width=6.4in]{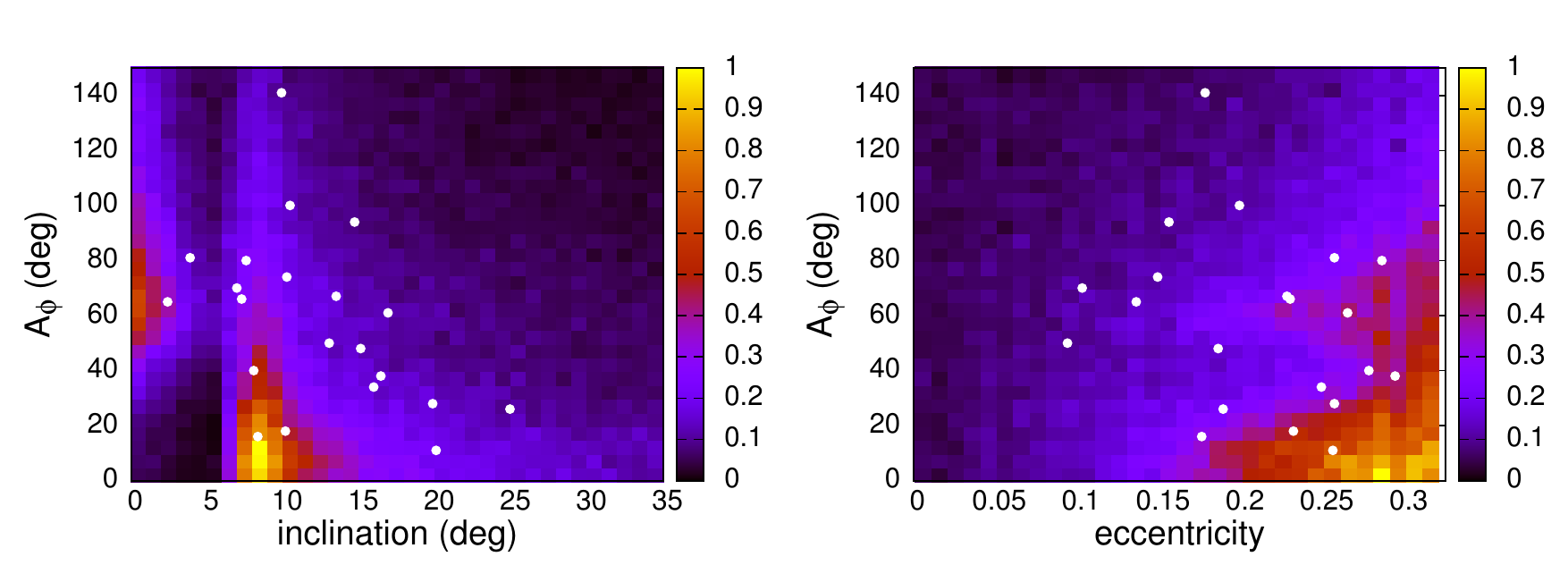}
\caption{Relative visibility (color coded) of $i-A_{\phi}$ and  $e-A_{\phi}$ plutino phase space for the OSSOS 13AO and 13AE blocks assuming uniform underlying distributions in orbital elements and an exponential $H$ magnitude distribution with a slope $\alpha=0.9$.  The white dots show the OSSOS detections.  The fact that the real detections do not cluster in the regions of high sensitivity simply indicates that the peaks of the intrinsic distributions lie at different values and that (unsurprisingly) a uniform underlying distribution for the population does not match the observations.}\label{f:32vis}
\end{figure}

Figure~\ref{f:32vis} shows how the sensitivity of the 13AO and 13AE blocks to plutinos varies in $e-A_{\phi}$ and $i-A_{\phi}$ phase space with the actual detections over-plotted in white; the relative visibilities are calculated using the survey simulator to simulate detections from a 3:2 population with uniform underlying distributions in the displayed ranges of $e$, $i$, and $A_{\phi}$ and a single exponential $H$ distribution with a slope $\alpha=0.9$ (see Section \ref{ss:32Hd}).   Of note in this figure is the survey's sensitivity to moderately inclined, low-$A_{\phi}$ plutinos due to the placement of the 13AO block near the libration center of the resonance and $\approx$$7^\circ$ above the invariable plane. The low-$A_{\phi}$ plutino phase space was not well explored by CFEPS, and G12 found that the $A_{\phi}$ distribution of the plutinos could be acceptably modeled with no $A_{\phi}<20^{\circ}$ component; this was also consistent with the scarcity of observed plutinos in the Minor Planet Center database with $A_{\phi}<20^{\circ}$ (\citealt{Lykawka2007} report one such object).  OSSOS has detected 3 moderately inclined plutinos with $A_{\phi}<20^{\circ}$ in 13AO block, showing that the low-$A_{\phi}$ part of the resonance is populated; we expect to further constrain the low-inclination, low-$A_{\phi}$ populations in an upcoming ecliptic block \citep[15BD, see ][]{Bannister2015} pointed near the center of the 3:2 resonance on the other side of Neptune.

\subsection{Plutino $i$ and $A_{\phi}$ distributions}\label{ss:32iaphi}

\begin{figure}[htpb]
   \centering
   \includegraphics[width=4.2in]{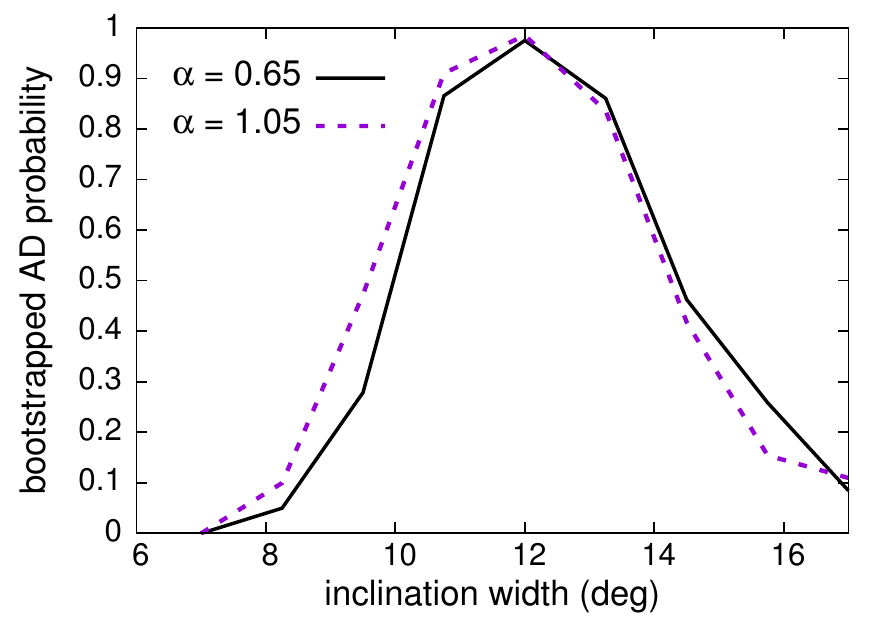}
\caption{The bootstrapped AD probability of various values of the inclination width $\sigma_i$ for two different $H$ distributions.  The rejectable range of $\sigma_i$ (AD probability below 0.05) does not change much when comparing two very different $H$ distributions. }\label{f:nocouple}
\end{figure}

We use the 21 OSSOS detections and the survey simulator to constrain acceptable models for the intrinsic plutino $i$ and $A_{\phi}$ distributions, as described in Section~\ref{ss:ssmod} and Appendix \ref{a:stats}. We model the inclination distribution using equation~\ref{eq:i}.  The AD test (Appendix \ref{sec-AD}) identifies an acceptable match between the inclinations of synthetic and real OSSOS detections  for $8^{\circ}\leq\sigma_i\leq 21^{\circ}$ at the 95\% confidence level.  This range is consistent with previous observational estimates of the plutino inclination width: $\sigma_i=8-13^{\circ}$ \citep[1-sigma confidence range]{Brown2001}, $\sigma_i~=~9-13^{\circ}$ \citep[1-sigma confidence range]{Gulbis2010}, $\sigma_i=12-24^{\circ}$ (G12, 95\% confidence range), and $\sigma_i=11-21^{\circ}$ \citep[95\% confidence range]{Alexandersen2015}.   We use a maximum likelihood approach (Appendix~\ref{sec-maxlike}) to determine a best-fit value of $\sigma_i=12^{\circ}$ for equation~\ref{eq:i}, although the probability distribution is quite flat in the range $\sigma_i=10-13^{\circ}$.  We also tested the acceptability of a Gaussian inclination distribution of the form
\begin{equation} \label{eq:i-offset-gaussian}
N(i) \propto \sin(i) \,exp{\left(-\frac{(i-i_c)^2}{2\sigma_i\,^2}\right)},
\end{equation}
which was used by \citet{Gulbis2010}.  Using the AD test, equation~\ref{eq:i-offset-gaussian} is a non-rejectable model for the plutino inclination distribution at 95\% confidence for $i_c<12^{\circ}$ with $\sigma_i$ ranging from $5-8^{\circ}$ at $i_c=12^{\circ}$.  However, a maximum likelihood comparison shows that based on the OSSOS detections an offset Gaussian (equation~\ref{eq:i-offset-gaussian}) is not a better description of the plutino inclination distribution than one centered on $0^{\circ}$ (equation~\ref{eq:i}), so we confine ourselves to the single parameter model. These results depend only weakly on the values for other model parameters, justifying our independent modeling of the $i$ distribution.  Figure \ref{f:nocouple} displays, as an example, the lack of coupling between the inclination and absolute magnitude distributions; we show that the bootstrapped AD probability for a range of $\sigma_i$ values does not significantly change for $H$ distributions with slopes $\alpha=0.65$ and $\alpha=1.05$ (values near the extreme ends of the 95\% confidence limits for $\alpha$ that we find in Section~\ref{ss:32Hd}). 

As discussed in Section~\ref{ss:ssmod}, we model the libration amplitude distribution as a triangle starting at a lower limit, $A_{\phi,min}$, peaking at $A_{\phi,c}$, and returning to zero at a maximum, $A_{\phi,max}$.  The OSSOS plutinos  have $A_{\phi}$ in the range $\sim10-140^{\circ}$, which constrains our choice of $A_{\phi,min}$ and $A_{\phi, max}$.  We ran a suite of models through the Survey simulator with $A_{\phi,min}<10^{\circ}$, $A_{\phi,max} = 140-170^{\circ}$, and $A_{\phi, c} = 20-120^{\circ}$.  We find that  $A_{\phi,min} = 0$, $A_{\phi,max} = 155^{\circ}$, and $A_{\phi,c} = 75^{\circ}$ provides the best match to the observed libration amplitudes (maximum likelihood), although the probability distribution in these parameters is quite flat.  Using the AD test, we cannot rule out  any values of $A_{\phi,min}$ or $A_{\phi,max}$ in our tested ranges.  At 95\% confidence we can constrain $A_{\phi,c}$ to be in the range $30-110^{\circ}$.  This range in $A_{\phi,c}$, although wide, represents a rigorous constraint on the libration amplitude distribution; detections from the remaining six OSSOS blocks should substantially improve this constraint.  Though multiple emplacement mechanisms could produce a libration amplitude distribution with multiple components, a single component model provides an acceptable fit to current data.

 We find a distribution of $A_{\phi}$ that, though mostly consistent with results of CFEPS (G12), contains additional objects with lower libration amplitudes than previously reported.   Future OSSOS blocks will provide additional 3:2 detections that will further constrain the $A_{\phi}$ distribution.

\subsection{Plutino $H$ and $e$ distributions}\label{ss:32Hd}

\begin{figure}[htpb]
   \centering
   \includegraphics[width=6.2in]{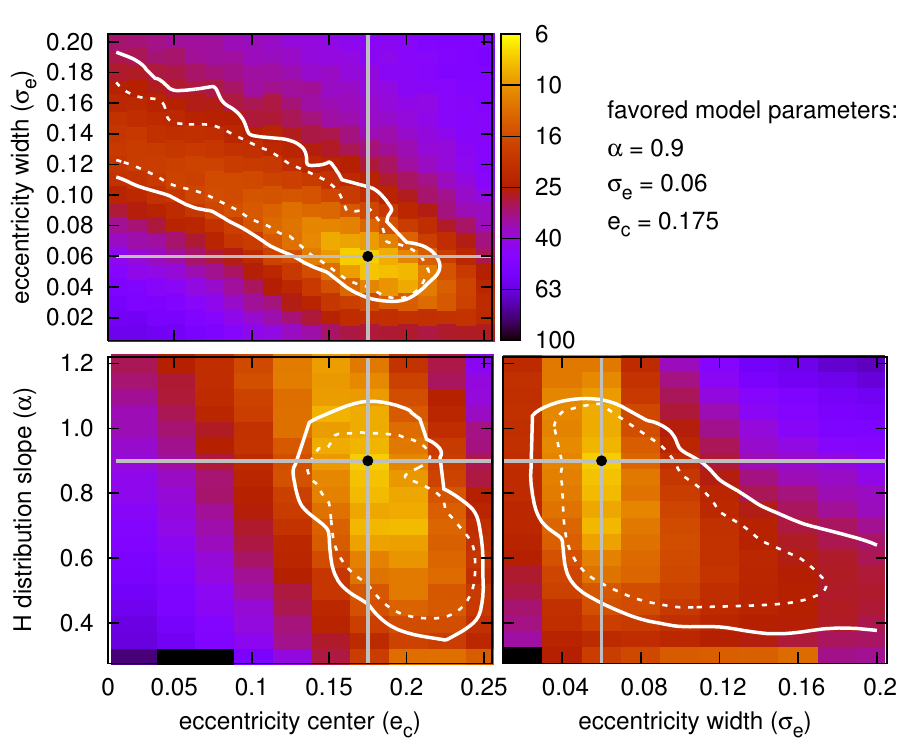}
\caption{Color maps: goodness of fit for various plutino model parameters as measured by a summed $\chi^2$ statistic for the $e$, $H$, and $d$ distributions. Lines: Rejected parameter values using the summed AD statistic for the $e$, $H$, and $d$ distributions at the 99\% confidence level (solid white curves) and the 95\% confidence level (dashed white curves).  Our favored model parameters (based on minimizing the summed $\chi^2$ statistic) are shown by the black dots.  Each panel is a 2-dimensional cut in our 3-dimensional parameter space search.  For each panel, we fix one parameter at its favored value and show the goodness of fit map for the other two parameters (for example, in the top panel, $\alpha$ is fixed at 0.9 to show the allowed range in $\sigma_e$ and $e_c$ for that value of $\alpha$). }\label{f:32edhmap}
\end{figure}

We ran a suite of survey simulations for plutino populations with a wide range of parameters for the eccentricity and $H$ distributions described by equations~\ref{eq:e} and~\ref{eq:spl}, respectively.  Because detection biases couple these distributions (Section \ref{s:bias}), we model them together. Our results are presented in Figure~\ref{f:32edhmap}.  The best-fit model, as measured by our summed chi-squared statistic (Appendix \ref{sec-CS}), is $\alpha=0.9$, $e_c = 0.175$ and $\sigma_e=0.06$, in agreement with the G12 results and derived from an observational sample that is completely independent from CFEPS.

The 21 OSSOS plutinos are acceptably modeled by a single exponential in $H$ with a slope $\alpha = 0.9^{+0.2}_{-0.4}$.  This is somewhat surprising given that previous surveys have shown that the dynamically excited TNO populations are not well-modeled by a single exponential. Recently \citet{Fraser2014} found that these populations can be modeled by a broken exponential $H$ distribution with a bright-end slope $\alpha=0.9$ that breaks to a faint-end slope $\alpha\sim0.2$ at $H_r(break)\sim8$.  \citet{Shankman2013} and \citet{Shankman2016} find that the scattering population shows evidence of a divot (a deficit of objects rather than a simple change in slope) in the $H$ distribution near $H_g\sim9$, corresponding to $H_r\sim8.4$.  \citet{Alexandersen2015} rejects a single exponential $H$ distribution for the plutinos, finding evidence for either a divot near $H_r\sim8.5$ or a break to a shallow slope at $H_r<8$.  Based on just the OSSOS sample, we cannot rule out a single exponential despite being sensitive to plutinos with $H_r>8$ where the divot or change in slope has been proposed.   

\begin{figure}[htpb]
   \centering
   \includegraphics[width=6.2in]{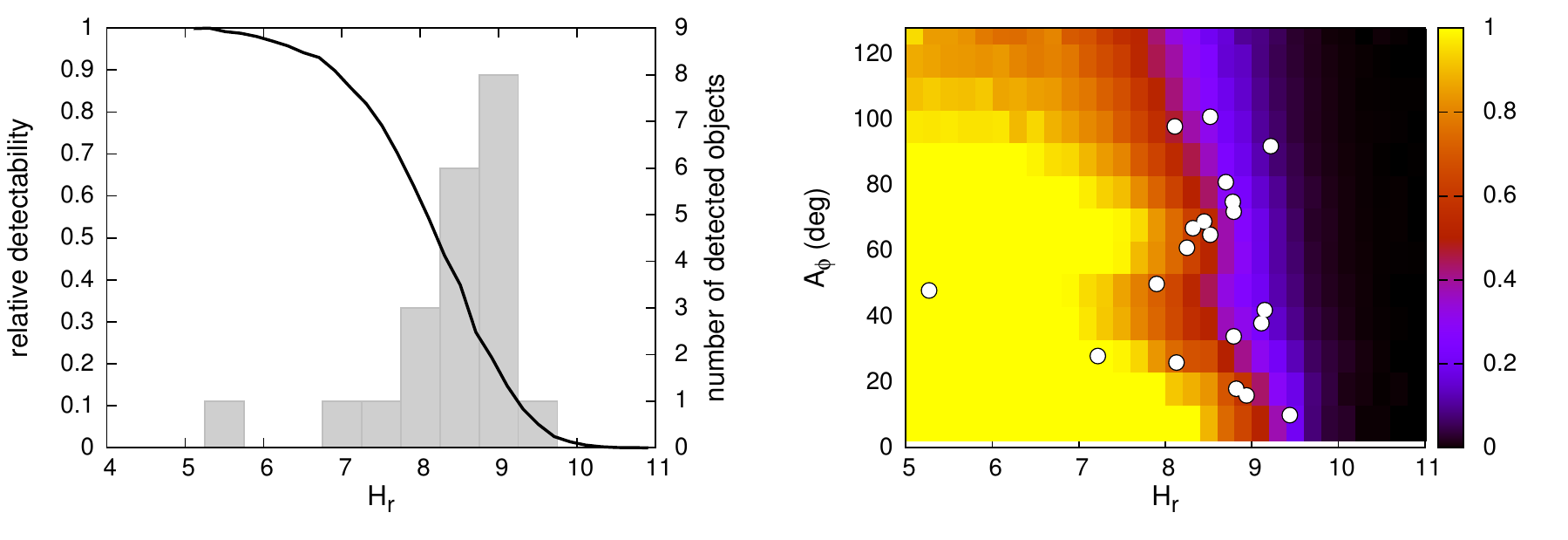}
\caption{Left panel: the black line shows the relative visibility of plutinos as a function of $H$ for the OSSOS 13AO and 13AE blocks assuming uniform underlying distributions. The gray histogram shows the actual number of detected plutinos as a function of $H$.  Right panel: color coded relative visibility of plutinos as a function of both $H$ and $A_\phi$ for the OSSOS 13AO and 13AE blocks assuming uniform underlying distributions.  The white dots show the OSSOS plutino detections.  The sensitivity to 60 degree libration amplitude is due to the location of 13AE block, which favors detection of plutinos with $A_{\phi}$ somewhat larger than $40^{\circ}$.}\label{f:32hvis}
\end{figure}

To examine the conflicting conclusions between OSSOS and the \citet{Alexandersen2015} results about the possibility of a single exponential all the way down to $H_r=9.2$, we generated 100 sets of 21 synthetic OSSOS detections for \citet{Alexandersen2015}'s preferred divot model.  We then tested how many of these 100 synthetic `observed' data sets would be able to reject our best-fit single exponential $H$ distribution.  We find that if the real plutinos follow \citet{Alexandersen2015}'s nominal divot distribution, a sample of 21 detected in the two OSSOS blocks would reject a single exponential $\sim80\%$ of the time.  So while we find no evidence of a transition in the OSSOS sample, this could just be due to our small sample size.  We note, however, that the placement of the OSSOS blocks means we were most sensitive to large-$H$ objects with low libration amplitudes, which differs from \citet{Alexandersen2015}'s survey. Figure~\ref{f:32hvis} shows the detectability of plutinos in the 13AO and 13AE blocks as a function of $H_r$ and of $A_{\phi}$ and $H_r$.  Many of the large-$H$ OSSOS objects have $A_{\phi}<40^{\circ}$, a previously sparsely observed part of the resonance's phase space.  It would be very interesting if the low $A_{\phi}$ plutinos have a different $H$ distribution than the larger $A_{\phi}$ plutinos; \citet{Lykawka2007} found some evidence for this in their analysis of the observed plutinos.  Different dynamical capture mechanisms populate different parts of the resonance, so it is not impossible that the low and high $A_{\phi}$ plutinos were captured from different parts of the primordial TNO population.  Better statistics afforded by the upcoming OSSOS blocks will further test this idea. 

\subsection{Plutino Kozai fraction}\label{ss:32kz}

Finally, we model the fraction of Plutinos that are also in the Kozai resonance.  Our dataset of 21 plutinos contains 5 Kozai oscillators.  Within the survey simulator, these objects are generated separately from the other plutinos because they occupy a distinct phase space within the resonance.  To account for this we follow the procedure outlined in G12 and \citet{Lawler2013} which uses an approximate Kozai resonant Hamiltonian \citep{Wan2007} to select values of $e$, $i$, and $\omega$ that correspond to Kozai libration of various amplitudes within the resonance.  A Kozai plutino's $H$ and $A_{\phi}$ are selected the same way as for the non-Kozai plutinos (we assume that Kozai and non-Kozai plutinos share a single libration amplitude distribution, which is sufficient to model current data).  Our procedure for choosing the other orbital parameters for Kozai plutinos is described in Appendix \ref{as:32}.

We ran a suite of survey simulations varying the intrinsic Kozai fraction ($f_{koz}$) from 0-1 to determine the probability of detecting 5 Kozai plutinos in an sample of 21 plutino detections for each value of $f_{koz}$.  To reproduce the 5 OSSOS 3:2 Kozai plutinos more than 5\% of the time, we find that the Kozai fraction must be $0.08-0.35$ ($f_{koz}=0.05-0.45$ at 99\% confidence).  An intrinsic Kozai fraction of 0.2 has the highest probability of reproducing the OSSOS detections.  This is in reasonable agreement with the $f_{koz}=0.1$ ($<$0.33 at 95\% confidence) determined by CFEPS  (G12).  As discussed in \citet{Lawler2013}, different resonant capture scenarios predict different values for $f_{koz}$; the first two OSSOS blocks have already narrowed the range of allowable $f_{koz}$ compared to the CFEPS results, and we expect the future blocks to provide an even better determination of the intrinsic Kozai fraction.

\subsection{Plutino population estimate and summary}\label{ss:32sum}

\begin{figure}[htpb]
   \centering
   \includegraphics[width=6.2in]{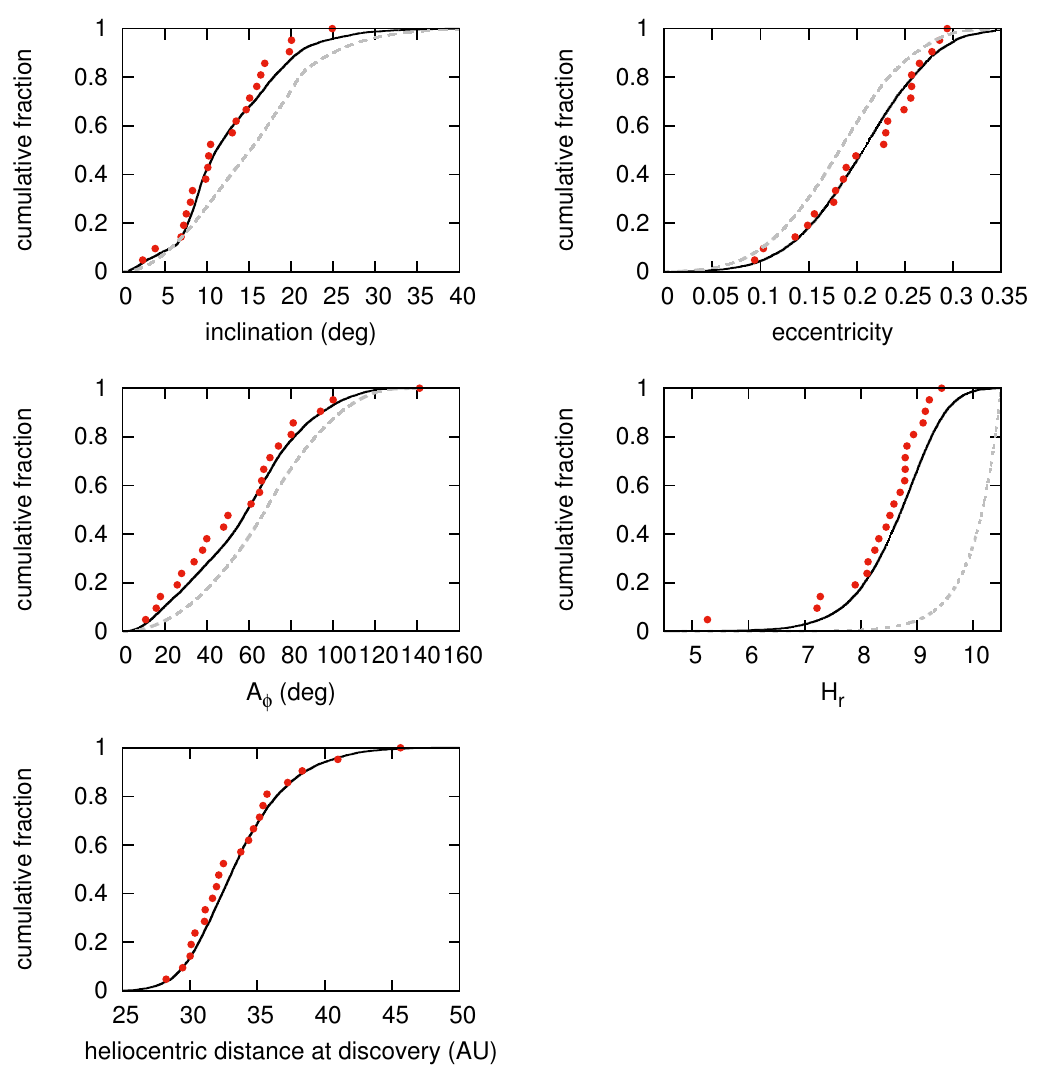}
\caption{Cumulative 1-d distributions in $i$, $e$, $A_{\phi}$, $H_r$, and distance at discovery for the observed 13AO and 13AE block plutinos (red dots), the intrinsic plutino population for our nominal plutino model (gray dashed lines) and for the synthetic detections from our nominal plutino model (black lines). The differences between the intrinsic models and the synthetic detections show the effects of the observational biases.}\label{f:32cdf}

\end{figure}

Our nominal best-fit values for the parameters in our plutino model are $\alpha=0.9$, $e_c = 0.175$, $\sigma_e=0.06$, $\sigma_i=12^{\circ}$, $f_{koz} = 0.2$, and a triangular $A_{\phi}$ distribution that goes from $0-155^{\circ}$ with a peak at $75^{\circ}$.  Figure~\ref{f:32cdf} shows this distribution compared to the actual OSSOS detections; there is generally good agreement in the one-dimensional distributions in $i$, $e$, $A_{\phi}$, $H_r$, and distance at discovery between the synthetic detections and the actual OSSOS detections. Using our best fit model, we estimate that the 3:2 resonance contains a population of $8000^{+4700}_{-4000}$ objects with $H_r < 8.66$ (see Section \ref{s:pop} for more details).  The independent OSSOS data sample yields best-fit orbital parameters and a total population estimate for the plutinos that are in good agreement with the CFEPS results (G12). 

\section{The surprisingly populous 5:2 resonance}\label{s:52}

\begin{figure}[htpb]
   \centering
      \includegraphics[width=6.2in]{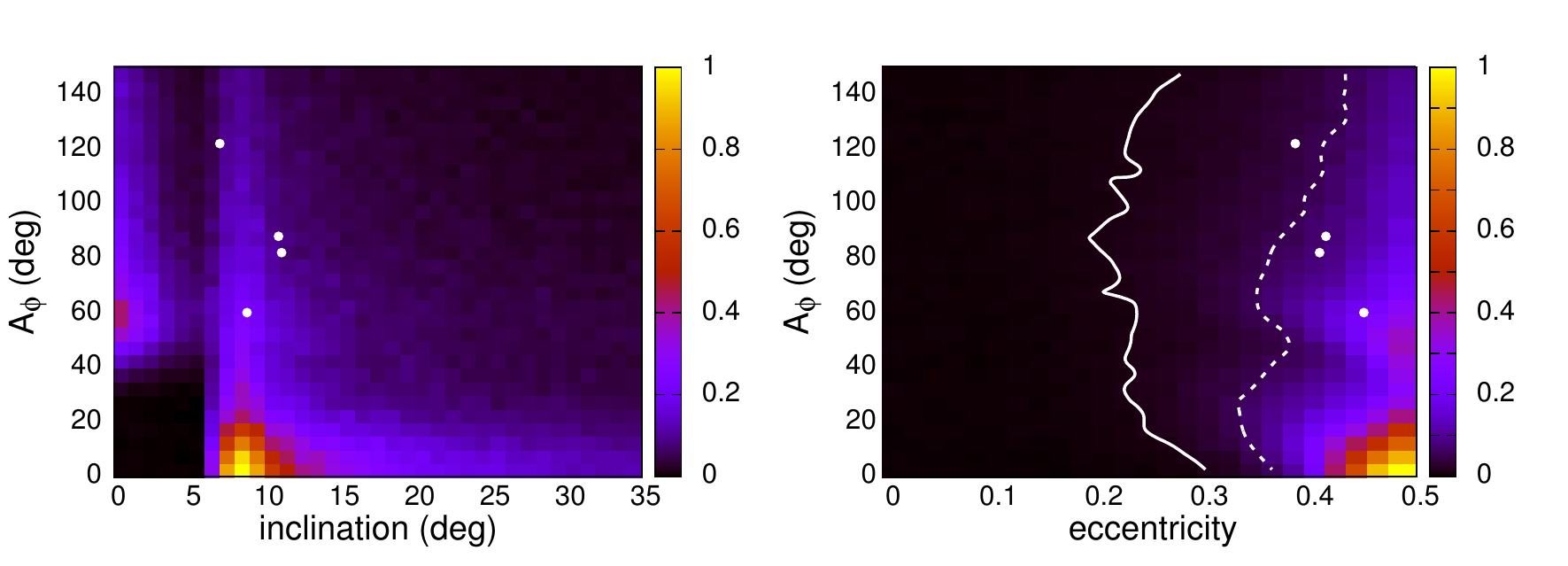}
      \caption{Relative visibility (color coded) of $i-A_{\phi}$ and $e-A_{\phi}$  5:2 phase space for the OSSOS 13AO and 13AE blocks assuming uniform underlying distributions.  The white dots show the OSSOS detections. In the right panel, the solid and dashed lines show the eccentricities below which visibility drops to $<1$\% and $<5$\% respectively for an $H$ distribution with $\alpha=0.9$. As in Figure~\ref{f:32vis}, the fact that the real detections do not cluster in the regions of high sensitivity simply indicates that a uniform underlying distribution in $e, i$, and $A_{\phi}$ does not match the observations.}\label{f:52vis}
\end{figure}

One of the surprising results from CFEPS was that the population of the 5:2 resonance was found to be nearly as large as the population of the 3:2 resonance (G12).  This is unexpected because planetary migration models do not predict efficient capture into the 5:2 resonance \citep[e.g.,][]{Chiang2002} and capture following dynamical instability \citep[e.g.,][]{Levison2008} likewise predicts a smaller 5:2 population relative to the 3:2.  So far, OSSOS has detected 4 objects in the 5:2 resonance at $a=55.5$ AU.  Given that the libration behavior of 5:2 resonant objects is similar to that of the 3:2, where objects at exact resonance come to perihelion at the ortho-Neptune points, the 13AO and 13AE blocks show a similar visibility profile for the 5:2 resonance (Figure~\ref{f:52vis}) as for the plutinos (Figure~\ref{f:32vis}).  The major difference between these two resonances is the much lower sensitivity to low-eccentricity 5:2 objects because it is a more distant resonance.  The right panel of Figure~\ref{f:52vis} shows contour lines in eccentricity below which the probability of observing an object from an eccentricity distribution uniform in the range $0-0.5$ drops below 5\% and 1\% assuming an underlying $H$ distribution with a slope $\alpha=0.9$; we don't expect a uniform eccentricity distribution, but this does demonstrate that OSSOS is not particularly sensitive to 5:2 objects with $e<0.3$.  

We use a parameterized orbital model for the 5:2 resonance identical to that for the non-Kozai plutinos.  We ran a suite of survey simulations to place limits on the parameterized $i$, $e$, and $H$ distributions.  Given the small number of detections, we used a single, triangular $A_{\phi}$ distribution that ranged from $0-140^{\circ}$ with a peak at $75^{\circ}$; this provided a statistically adequate representation of the OSSOS 5:2 detections and is similar to the $A_{\phi}$ distribution used in G12 for this population.  The upper limit for libration in the 5:2 resonance (from both observations and numerical integrations) appears to be $A_{\phi}\sim155^{\circ}$ \citep{Lykawka2007,Lykawka2007s}, but the extension of the $A_{\phi}$ distribution above $140^{\circ}$ is not necessary to describe the OSSOS 5:2 detections; with the future OSSOS blocks, we expect more 5:2 detections and will explore the upper limit for the $A_{\phi}$ distribution. 

We find the inclination distribution can be modeled using equation~\ref{eq:i} with a most-probable (maximum likelihood) width of $\sigma_i=10^{\circ}$.  At 95\% confidence using the AD statistic, the width ranges from $6-20^{\circ}$ in agreement with the width of $\sigma_i = 15^{\circ}$ for the 5:2 from G12.  

\begin{figure}[htpb]
   \centering
   \includegraphics[width=4.2in]{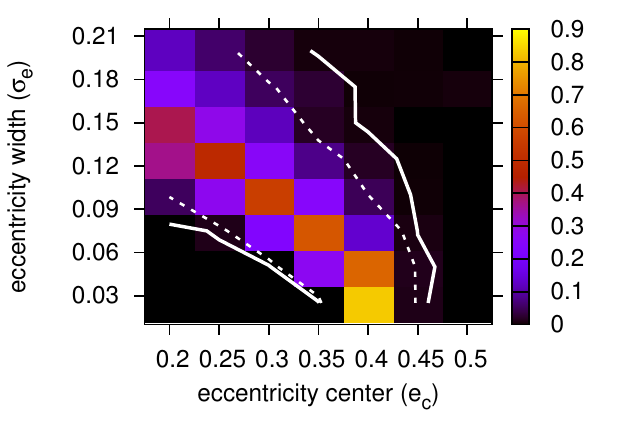}
\caption{AD rejectability of a 5:2 eccentricity distribution with width $\sigma_e$ and center $e_c$ assuming an underlying $H_r$ distribution of slope 0.9.  The lines indicate values rejectable at 99\% (solid white) and 95\% (dashed white) confidence.}\label{f:52ewec}
\end{figure}

As discussed for the plutinos, the $e$ and $H$ distributions can't be determined independently from each other.  We find that a single exponential is an adequate model for the 4 OSSOS detections; reasonable eccentricity distributions can provide acceptable matches for the observed $e$, $H$, and heliocentric distance at discovery for slopes in the range $0.6 < \alpha < 1.1$ with no strongly preferred value (based on a summed chi-square statistic).   With only 4 detections, it is not surprising that we don't have a strong constraint on $\alpha$.  Assuming the 5:2 population has the same $H$ distribution as the plutinos, we can constrain the allowable range of eccentricity distribution parameters (equation~\ref{eq:e}).  Figure~\ref{f:52ewec} shows the significance levels of the summed AD statistic for the $d$, $e$, and $H$ distributions of the 4 OSSOS 5:2 detections compared to simulated detections for a range of $e_c$ and $\sigma_e$ values.  Because the observed 5:2 objects have a narrow range in $e$ of $0.39-0.45$, the least-rejectable eccentricity distribution has $e_c=0.4$ and $\sigma_e=0.025$.  However this is not likely to be a good representation of the true 5:2 eccentricity distribution; there are 5:2 objects with $e\sim0.3$ in the MPC database \citep[also listed in][]{Gladman2008,Lykawka2007,Adams2014} which invalidates such a strongly peaked $e$ distribution centered at $e_c=0.4$.  As Figure~\ref{f:52ewec} shows, the OSSOS detections do not rule out $e$-distributions with smaller $e_c$ and larger $\sigma_e$, a result that is consistent with the findings of G12; however, the insensitivity of the OSSOS 2013AO/E blocks to 5:2 objects with $e\lesssim0.3$ makes this distribution difficult to constrain.  If we limit our model to $e>0.35$, we find that the OSSOS observations can be adequately reproduced by a uniform eccentricity distribution in the range $e=0.35-0.45$.  We use this restricted $e$ range to model the total intrinsic population of the 5:2 with the understanding that this makes our population estimate a lower limit because we know that the $e<0.35$ region is occupied.  For our best-fit model applied to OSSOS data alone, we find that the 5:2 resonance contains $5700^{+7300}_{-4000}$ objects with $H_r < 8.66$ and $e >0.35$ (see Section~\ref{s:pop} for more details).

\section{New constraints on the symmetric to asymmetric ratio for the 2:1 resonance}\label{s:21}

The 2:1 is the strongest of the n:1 resonances.  In the 2:1, symmetric librators have a resonant angle $\phi$ (see Section \ref{s:methods}) which, like that for all 3:2  objects, librates about $180^{\circ}$.  Asymmetric librators instead librate about a center near $\phi\sim60-100^{\circ}$ or $\phi\sim260-300^{\circ}$.    \citet{Nesvorny2001} studied the current dynamics of the 2:1 resonance, determining  how the libration centers and amplitudes change with eccentricity and how the stability of the resonance is affected by inclination.  \citet{Tiscareno2009} also studied the stability of 2:1 phase space.  Determining how the current 2:1 resonant objects are split between the symmetric, leading asymmetric, and trailing asymmetric libration islands is of particular interest for determining how this resonance became populated; \citet{Chiang2002} and \citet{Murray-Clay2005} demonstrated that Neptune's migration speed affects the probability of capture into the leading or trailing asymmetric libration centers, with higher speed migration favoring the trailing island. In this section we describe how we use the first two OSSOS blocks to constrain the fraction of symmetric 2:1 librators.  We then use the combined OSSOS and CFEPS observations to provide a well-characterized constraint on the trailing-to-leading ratio in the 2:1 resonance; as we discuss later, the combined data set is used for this constraint because the first two OSSOS blocks were only sensitive to trailing 2:1 asymmetric librators.

Because of the more complicated phase space of the 2:1 resonance compared to the 3:2 or 5:2 resonances, we do not have a simple parameterized $A_{\phi}$ distribution for the 2:1.  Both the libration centers  and the allowable range of $A_{\phi}$ for the asymmetric islands are $e$-dependent.  We also only have 4 OSSOS detections, so an overly complicated model is not warranted.  We base our 2:1 model on the results of \citet{Nesvorny2001}, who published a plot of libration centers and maxiumum libration amplitudes as a function of $e$.  To generate a 2:1 population, we first decide if an object is symmetric or asymmetric.  If it is symmetric, we select $e$ from a uniform range $0.05-0.35$ and $A_{\phi}$ from a uniform range $135-165^{\circ}$; these ranges correspond to the regions of relatively stable libration found in theoretical and numerical experiments \citep{Nesvorny2001,Chiang2002,Tiscareno2009}.  For asymmetric librators, we select $e$ uniformly from $0.1-0.4$.  For the chosen value of $e$, we choose the libration center from \citet{Nesvorny2001} and then assign a libration amplitude uniformly from $0-A_{\phi, max}$.  The inclinations are randomly selected from a Gaussian inclination distribution described by equation~\ref{eq:i}.

From just the 4 OSSOS detections, we find that the above simplified model for the 2:1 resonance (only slightly modified from the CFEPS 2:1 model of G12) is consistent with the observations.  We find that the inclination distribution width must be $\sigma_i<8^{\circ}$ at 95\% confidence with a most-probable value of $4^{\circ}$, independently confirming G12's conclusion that the 2:1 population is significantly colder in inclination than either the 3:2 or the 5:2.  We note that there are a few observed 2:1 objects in the MPC database with inclinations in the range $\sim20-30^{\circ}$. Most of these high inclination 2:1 objects appear to be large amplitude symmetric librators (see for example Table 1 in \citealt{Lykawka2007}). \citet{Tiscareno2009} showed that high inclination symmetric librators are not stable on Gyr timescales; this perhaps indicates that these observed large inclination 2:1 objects (a population not yet detected by OSSOS) are only temporarily stuck to the 2:1 resonance rather than primordial members.  We will explore the possibility of a population of higher-inclination, temporary 2:1 objects in addition to the low-$i$ (presumably primordial) 2:1 population with future OSSOS observations.

Based on the fact that half of the OSSOS 2:1 objects are symmetric librators, we can place a weak limit on the intrinsic fraction of symmetric 2:1 objects, $f_{s}$.  For our parameterized model of the 2:1 resonance, we tested intrinsic symmetric fractions ranging from 0.05-0.95.  For each tested $f_{s}$ we can determine the probability of drawing 4 synthetic observed objects with a $f_{s,obs}\ge0.5$.   This probability allows us to rule out $f_{s}\le0.05$ at the 99\% confidence level and $f_{s}\le0.1$ and $f_s ge0.95$ at the 95\% confidence level.

\begin{figure}[htpb]
   \centering
   \includegraphics[width=6.2in]{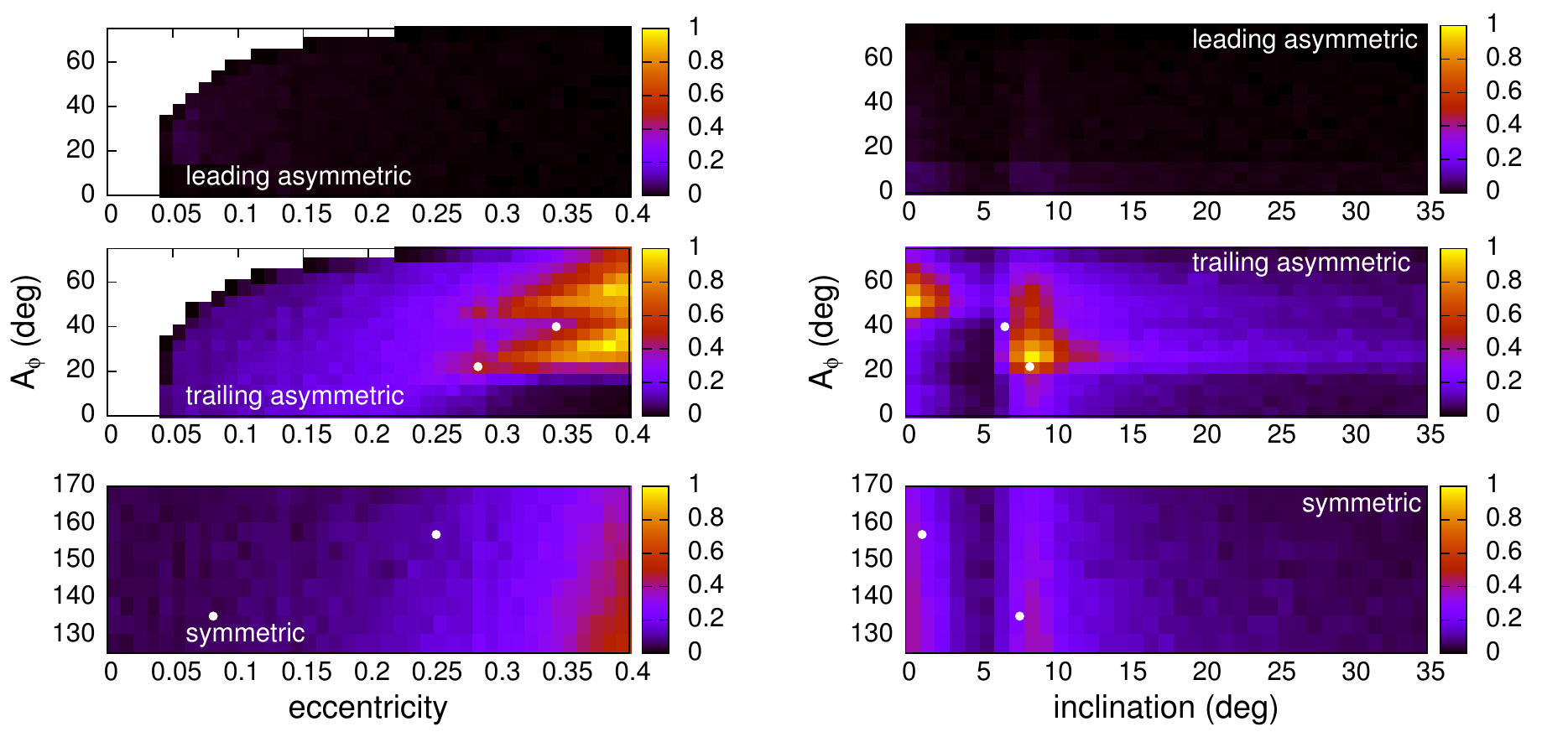}
\caption{Relative visibility (color coded) of $e-A_{\phi}$ and $i-A_{\phi}$ 2:1 phase space for the OSSOS 13AO and 13AE blocks assuming an even split between the leading asymmetric, trailing asymmetric, and symmetric libration centers as well as uniform $e$, $i$, and $A_{\phi}$ distributions within the resonant phase space of all three libration islands.  The white dots show the OSSOS detections.}\label{f:21vis}
\end{figure}

\begin{figure}[htpb]
   \centering
   \includegraphics[width=6.2in]{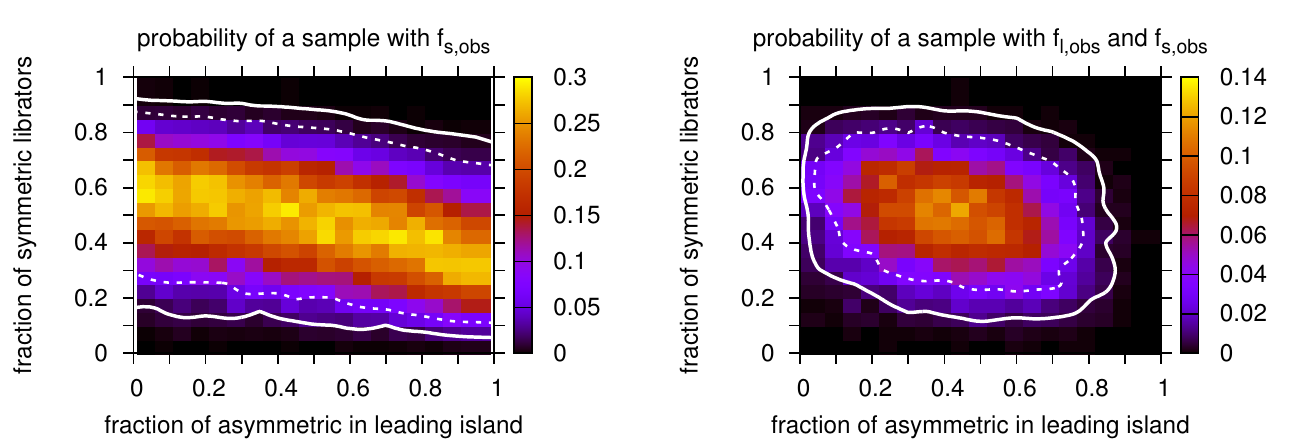}
\caption{Color maps: probability distributions comparing the simulated 2:1 detections to the combined CFEPS and OSSOS 2:1 detections.  Left: the probability of having three symmetric librators in a sample of nine 2:1 objects randomly drawn from the survey simulator's synthetic detections as a function of the simulated 2:1 population's intrinsic fraction of symmetric librators (y-axis) and intrinsic fraction of asymmetric 2:1 objects librating around the leading libration center (x-axis). Right: the probability of drawing a sample of nine 2:1 objects with three symmetric librators, five trailing asymmetric librators, and one leading asymmetric librators from the simulated detections.  In both panels, the rejected regions for the probability distributions are over plotted as solid white curves (99\% confidence level) and dashed white curves (95\% confidence level).}\label{f:21maps}
\end{figure}

To further constrain the allowable range of $f_s$, we repeat this calculation with the 9 combined CFEPS and OSSOS 2:1 detections while additionally considering the division of the asymmetric librators between the leading and trailing libration centers.  The two OSSOS blocks both point toward the trailing libration center, them fairly insensitive to the leading/trailing fraction.  This is evident in Figure~\ref{f:21vis}, which shows the relative visibility of all three libration islands in $e$-$A_{\phi}$ and $i$-$A_{\phi}$ phase space; the probability of detecting a leading asymmetric 2:1 object in the OSSOS 13AO or 13AE blocks is nearly 0.  Additional OSSOS blocks will cover the leading center, but for now we can use the CFEPS detections in addition to the OSSOS 13AO and 13AE block detections because CFEPS covered both libration centers (G12).  Of the 9 combined OSSOS and CFEPS 2:1 detections, 3 are symmetric librators and 6 asymmetric; 5 of the asymmetric detections are in the trailing libration island and 1 is in the leading island.  We ran a suite of OSSOS+CFEPS survey simulations for a wide range of intrinsic symmetric fractions,  $0.05<f_{s}<0.95$,  and a wide range of the intrinsic fraction of asymmetric librators in the leading libration center, $0<f_{lead}<0.95$.  The left panel of Figure~\ref{f:21maps} shows the probability of drawing a sample from the synthetic detections for each combination of $f_{s}$ and $f_{lead}$ that matches the observed symmetric fraction, $f_{s,obs}=1/3$; the right panel shows the probability of drawing a sample with $f_{s,obs}=1/3$ and $f_{lead,obs}=1/6$.   Using the combined OSSOS and CFEPS detections, we have the constraint that $0.1<f_{s}<0.9$ at the 99\% confidence level and $0.2<f_{s}<0.85$ at the 95\% confidence level; the fraction of asymmetric objects in the leading libration center is constrained to be  $f_{lead}<0.9$ at the 99\% confidence level and $0.05<f_{lead}<0.8$ at the 95\% confidence level.

To obtain a population estimate for the 2:1 resonance, we assume that the population is evenly split between symmetric and asymmetric librators and that the asymmetric librators are evenly split between the leading and trailing islands ($f_s=0.5$ and $f_{lead}=0.5$); we also assume the $H$ distribution has $\alpha=0.9$. Using this model and just the OSSOS data, the 2:1 resonance is estimated to contain $5200^{+9000}_{-4000}$ objects with $H_r<8.66$ (see Section \ref{s:pop} for more details).

\section{Population estimates}\label{s:pop}

\begin{figure}[htpb]
   \centering
   \includegraphics[width=4.2in]{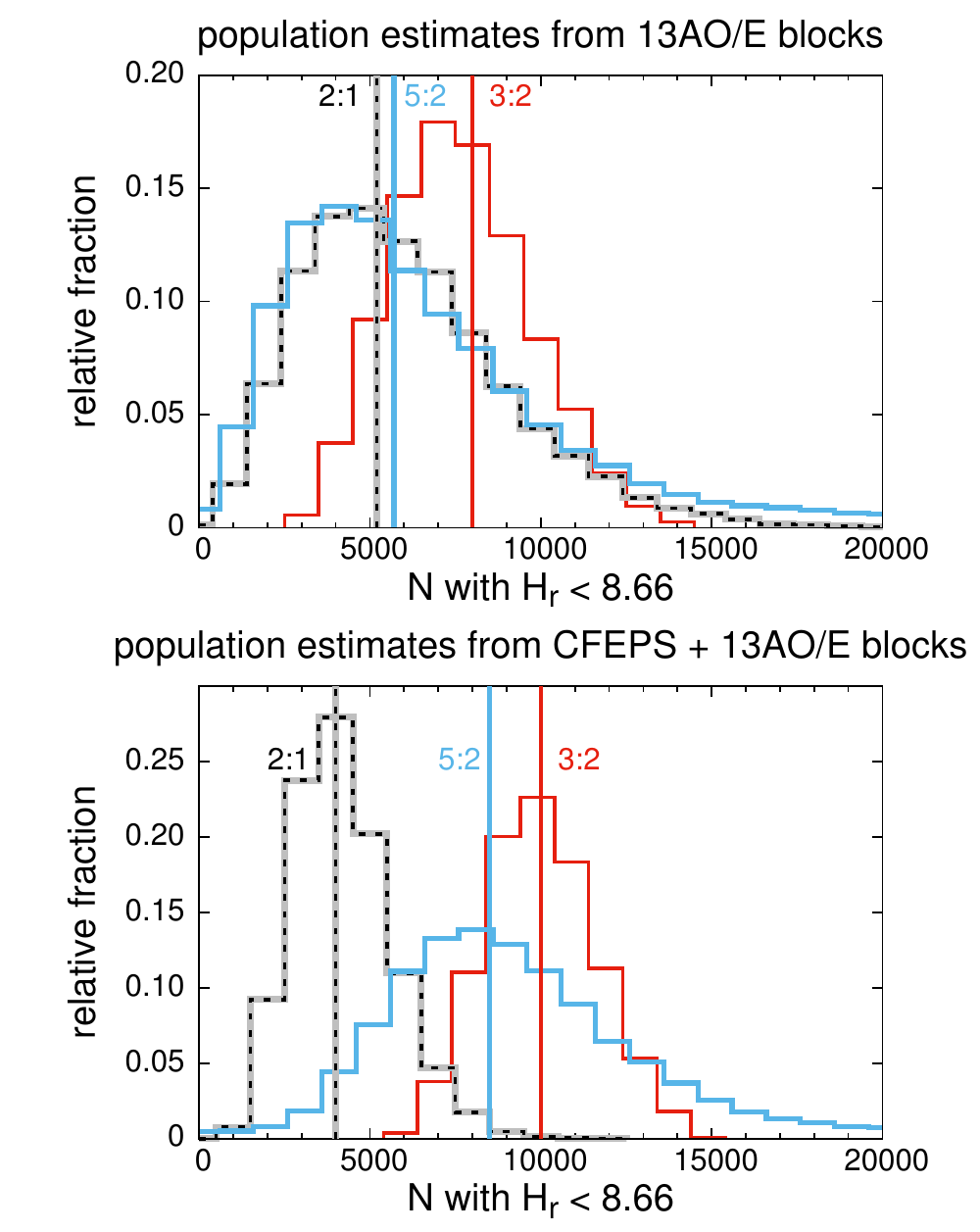}
\caption{Histogram of population estimates from 10000 survey simulator runs for our nominal 3:2, 5:2, and 2:1 population models.  The top panel shows the results for just the OSSOS detections for the 13AE and 13AO blocks.  The bottom panel shows the results for combining the two OSSOS blocks with CFEPS.}\label{f:pops}
\end{figure}

We have modeled the 3:2, 5:2, and 2:1 resonances based on the first set of OSSOS detections.  From this independent data set, we find that the orbital and $H$ distributions for these resonances are consistent with those found by CFEPS (G12).  Taking our nominal orbital models based on the OSSOS detections, we can construct population estimates for these three resonances.  To do this we run $10^4$ instances of the survey simulator for our orbital models of each resonance and determine how many objects with $H_r$ less than some limiting value must be generated to match the 21 plutino detections, the four 2:1 detections, and the four 5:2 detections in the 13AO and 13AE blocks.  To facilitate comparison to the population estimates from CFEPS, we choose a limiting magnitude $H_r=8.66$ to compare to their $H_g=9.16$.  This assumes the resonant populations have colors $g-r=0.5$, which matches the value (within photometric uncertainties) for the plutino population \citep{Alexandersen2015} based on the $g-r$ colors of the CFEPS 3:2 objects \citep{Petit2011};  there is an ongoing Gemini program \citep{Fraser2015} that will quantify the colors for the OSSOS $m_r < 23.5$ resonant objects, so in future analysis this color assumption might be improved. Our population estimates are listed in Table~\ref{t:pops} and shown in Figure~\ref{f:pops}.  All three population estimates overlap with the 95\% confidence bounds on the CFEPS (G12) estimates, although our median number of plutinos and our lower limit for the 5:2 population are both smaller than the CFEPS estimates and our 2:1 population is slightly larger.  Our 2:1 population estimate is much more uncertain that the CFEPS estimate despite the roughly equal numbers of OSSOS and CFEPS 2:1 detections; this is due to the restricted longitude range of the first two OSSOS blocks (both trailing Neptune) compared to the wider longitude ranges of the CFEPS observations.

\begin{deluxetable}{c l l r r}
\tabletypesize{\footnotesize}
\tablecolumns{5}
\tablewidth{0pt}
\tablecaption{Population Estimates}
\tablehead{\colhead{Res} & \colhead{$e$ distribution} & \colhead{$i$ distribution} & \colhead{13AO/E blocks} & \colhead{13AO/E + CFEPS} \\ \colhead{} & \colhead{} & \colhead{} & \colhead{$N(H_r<8.66)$} & \colhead{$N(H_r<8.66)$} }
\startdata
3:2 	& Eq.~\ref{eq:e}, $e_c = 0.175$, $\sigma_e = 0.06$ 		& Eq.~\ref{eq:i}, $\sigma_i = 12^{\circ}$ 	& $8000^{+4700}_{-4000}$ & $10000^{+3600}_{-3000}$ \\[5pt]
5:2 	& uniform $e = 0.35-0.45$ 						& Eq.~\ref{eq:i}, $\sigma_i = 11^{\circ}$	& $5700^{+7300}_{-4000}$ & $ 8500^{+7500}_{-4700}$\\[5pt]
2:1 & sym: uniform $e=0.05-0.35$						& Eq.~\ref{eq:i}, $\sigma_i = 4^{\circ}$ 	& $5200^{+9000}_{-4000}$ & $4000^{+2500}_{-2000} $\\
 	& asym: uniform $e=0.1-0.4$ 						&&&\\
\enddata
\tablecomments{Population estimates for the resonances with multiple secure OSSOS detections.  The population estimate is the median number of $H_r<8.66$ objects the survey simulator had to generate using our nominal models (described in Section~\ref{s:32}, \ref{s:52}, and~\ref{s:21}) to produce the observed number of detections with 95\% limits stated. The limit of $H_r=8.66$ is equivalent to the limit of $H_g = 9.16$ used in G12 assuming an average color for the resonant objects of $g-r = 0.5$.}
\label{t:pops}
\end{deluxetable}

\begin{figure}[htpb]
   \centering
   \includegraphics[width=6.0in]{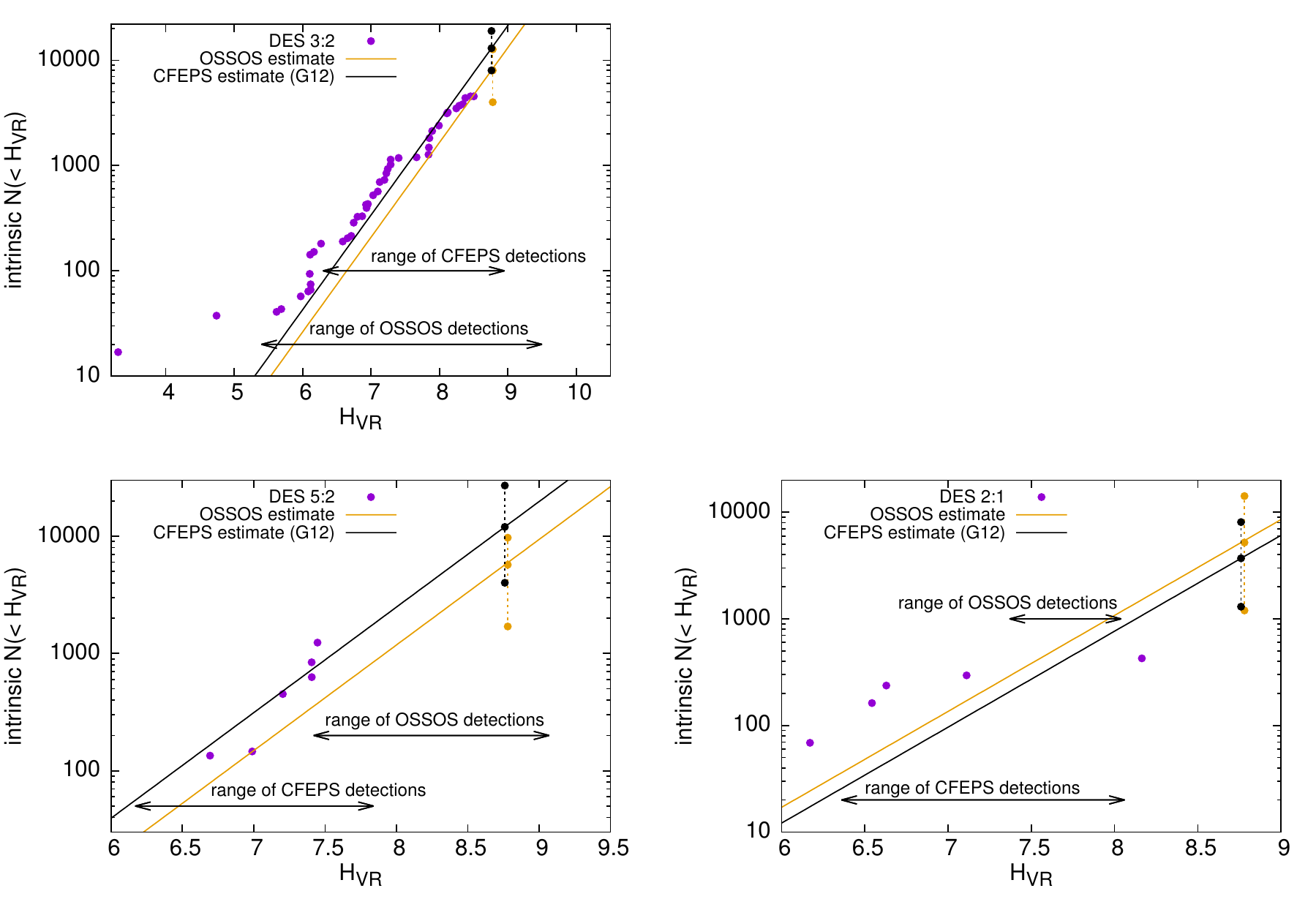}
\caption{Comparison of the DES \citep{Adams2014} 3:2, 5:2, and 2:1 population estimates (data taken from their Figure 1) to the population estimates for CFEPS (G12) and the first two OSSOS blocks (this paper), when shifted to the VR system.  The solid lines for CFEPS and OSSOS show our estimated best fit exponential H distributions with a slope $\alpha=0.9$ anchored at the $N(H_g<9.16)$ values from G12 for CFEPS and the $N(H_r<8.66)$ values from Table~\ref{t:pops}; the estimated 95\% confidence limits are shown as dashed lines for both CFEPS and OSSOS.  The arrows indicate the approximate range in $H_{VR}$ where CFEPS and OSSOS had detections for each resonance; each purple dot for the DES results is an individual detection and thus shows the DES observed range of $H_{VR}$. We assume color conversions of $g-VR=0.4$ and $g-r=0.5$.}\label{f:descompare}

\end{figure}

We can also compare our population estimates to those from the Deep Ecliptic Survey (DES) given in \citet{Adams2014}.  The DES observations were done in the VR filter, so we must assume a value of $VR - r$ in order to compare the population estimates.  \citet{Adams2014} assumed $g-VR = 0.1$ for comparing the DES population estimates to the CFEPS population estimates and found that the two sets of population estimates for the 3:2 population were discrepant.  However, we find that $g-VR = 0.4$ is a better estimate of the color conversion for the resonant objects based on a comparison of H values in the two color filters for resonant objects observed by both surveys (see Appendix~\ref{a:colors}).  Using the measured $g-VR=0.4$ color rather than $g-VR=0.1$ erases the discrepancy between the DES and CFEPS 3:2 population estimates reported by \citet{Adams2014}.  Figure~\ref{f:descompare} shows the data from Figure~1 in \citet{Adams2014} along with the resonant population estimates from Table~3 in G12 and Table~\ref{t:pops} of this work (taking $g-VR=0.4$ and $g-r=0.5$).  We find that the CFEPS and OSSOS population estimates for the 3:2, 5:2, and 2:1 resonances are in very good agreement with the intrinsically faint (large H) DES population estimate for the 3:2 resonance and overlap with the DES 2:1 and 5:2 population estimates.

Our median population estimates imply that intrinsic ratio of 3:2 / 5:2 / 2:1 objects is 1.5 / 1.1 / 1, compared to 3.5 / 3.2 / 1 from the CFEPS population estimates (G12) and 1.4 / 0.7 / 1 from the Deep Ecliptic Survey population estimates \citep{Adams2014} (note that because the color conversion is assumed to be the same for all the resonances, the DES population ratios are independent of the assumed $g-VR$ color).  Combining the OSSOS and CFEPS detections to obtain population estimates for our nominal resonance models (also listed in Table~\ref{t:pops} and shown in Figure~\ref{f:pops}), we find a ratio of 2.5 / 2 / 1. The uncertainties in our population estimates from just the OSSOS data are currently too large to conclusively determine whether the 5:2 is more populated than the 2:1 or as populated as the 3:2, but the OSSOS detections are consistent with a large population in the 5:2 resonance.  Additionally, we have used an artificially restricted eccentricity range for the 5:2 resonance due to our insensitivity to $e<0.35$ objects, so the real 5:2 population is likely to be larger than our estimate.

\section{Other resonances}\label{s:other}

In the OSSOS 13AO and 13AE blocks there are detections in nine other resonances: the 8:5, 18:11, 5:3, 16:9, 15:8, 7:3, 7:4, 13:5, and 11:4 resonances.  Of these detections, only the 7:4 and 7:3 detections are securely resonant as defined by the \citet{Gladman2008} classification scheme.  We integrated many clones of each insecure resonant detection to determine the probability that the objects are resonant; these probabilities are listed in Table~\ref{t:detections}.  Two of the insecure resonant detections have best-fit orbits that show libration of a mixed resonant argument.  OSSOS object o3o32 shows libration of the angle $\phi = 18\lambda_{tno} - 11 \lambda_N - 5 \varpi_{tno} - 2 \Omega_{tno}$ and object o3e49 shows libration of the angle  $\phi = 15\lambda_{tno} - 8 \lambda_N - 5 \varpi_{tno} - 2 \Omega_{tno}$.  

Single and/or insecure detections are not enough to characterize the structure of a resonance or provide a well-constrained population estimate, but we can check whether our single secure detections for the 7:4 and 7:3 resonances are consistent with the G12 models and population estimates for these resonances.  To test the 7:3 and 7:4 resonance models, we ran the G12 parameterized models through the OSSOS survey simulator to generate 10000 simulated detections for the 7:4 and 7:3.  In both cases the observed characteristics of the real OSSOS detections ($e$, $i$, $d$, $H_r$, and $A_{\phi}$) fall within the 95\% bounds of the synthetic detections, indicating that the G12 models are consistent with the OSSOS detections.  We also generated population estimates for these models of the 7:4 and 7:3 resonances.  For the 7:4 resonance the median population of objects with $H_r < 8.66$ is 1000 with a 95\% confidence range of $50-5000$ which agrees with the G12 95\% confidence estimate of $1000-7000$ objects with $H_g < 9.16$ (assuming a $g-r=0.5$).  For the 7:3 resonance, the median population of objects with $H_r < 8.66$ is 4000 with a 95\% confidence range of $100-20000$ again in agreement with the G12 95\% confidence estimate of $1000-12000$ objects with $H_g < 9.16$.  Testing of the other G12 resonance models and population estimates will be presented in future papers as more OSSOS blocks are completed and the orbits of the remaining insecure 13AE and 13AO resonant detections are improved by follow-up observations.

\section{Discussion and summary}\label{s:sum}

\begin{figure}[htpb]
   \centering
   \includegraphics[width=6in]{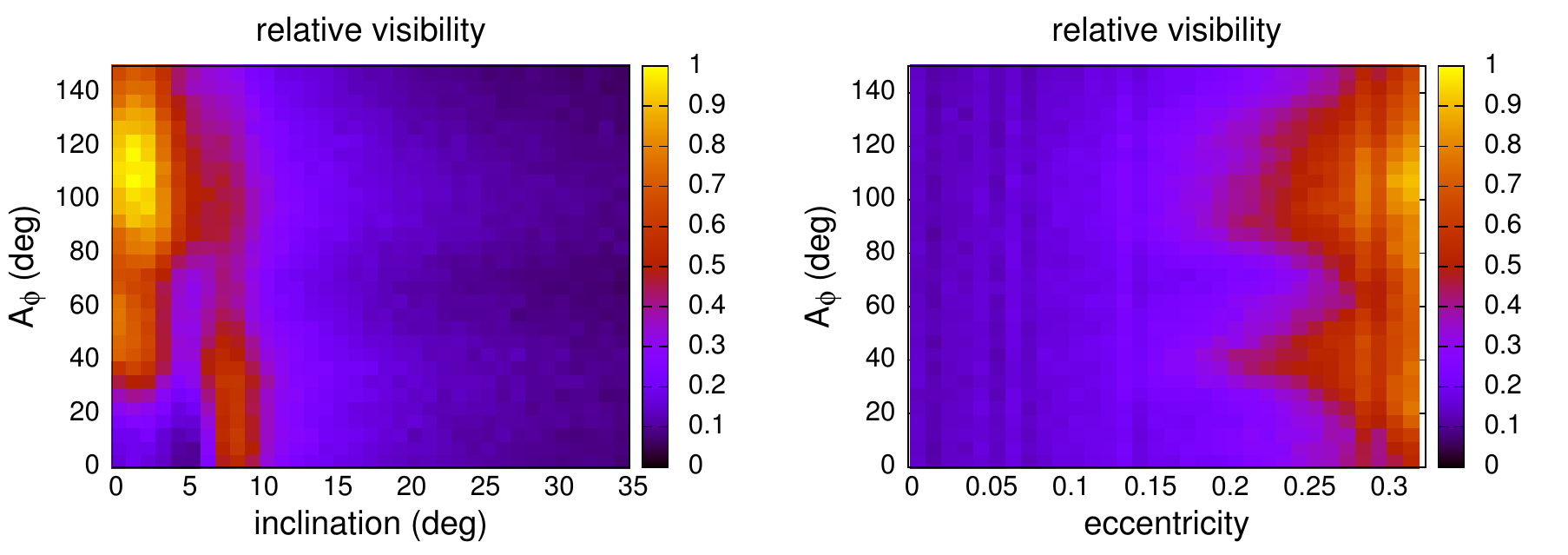}
\caption{Estimated visibility map for the 3:2 resonance in the full OSSOS survey assuming a uniform underlying distribution in $e$, $i$, and $A_{\phi}$ and a slope of $\alpha=0.9$ for the $H$ distribution (see Section~\ref{s:32} and Figure~\ref{f:32vis} for comparison).}\label{f:visfull32}
\end{figure}

\begin{figure}[htpb]
   \centering
   \includegraphics[width=6in]{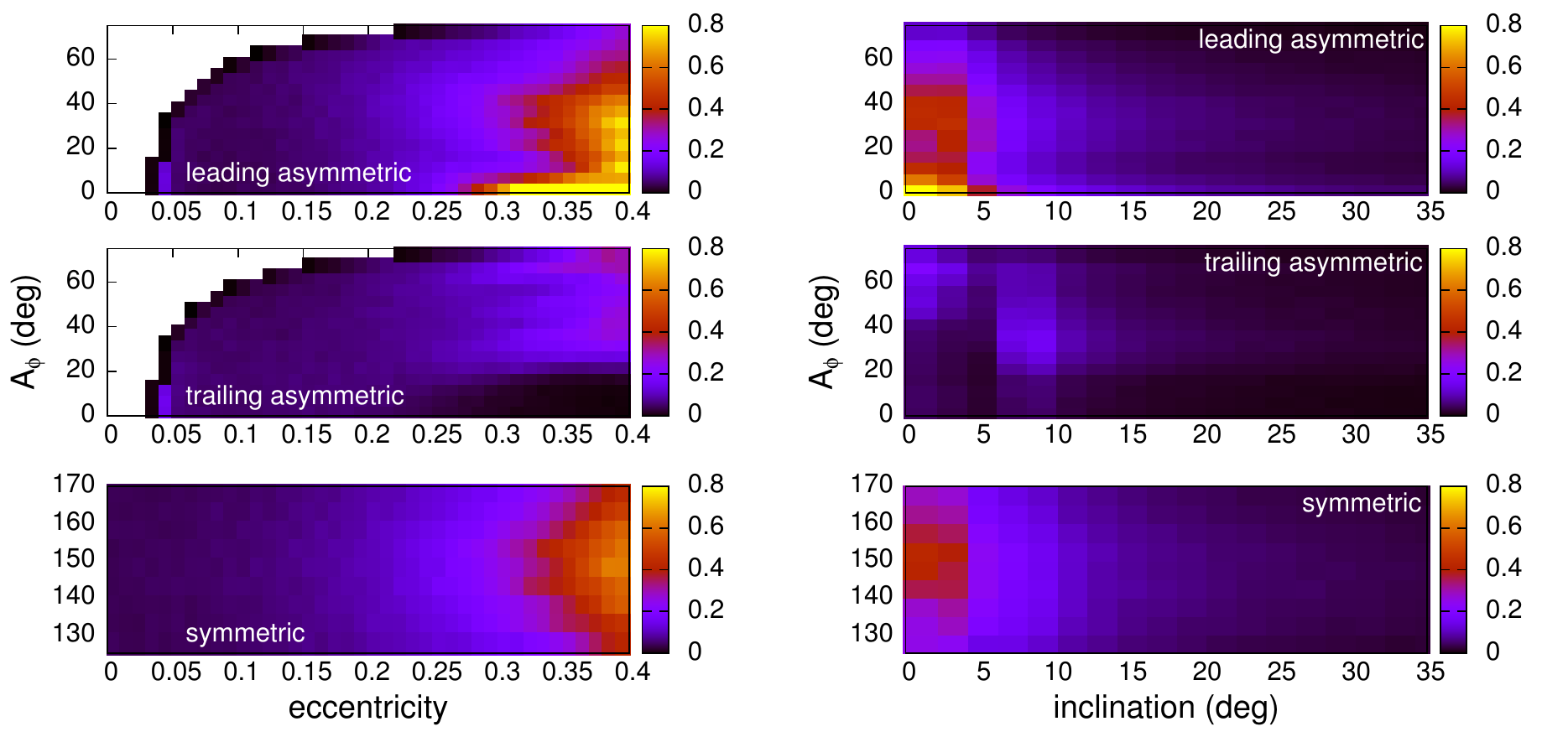}
\caption{Estimated visibility map for the 2:1 resonance in the full OSSOS survey assuming a uniform underlying distribution in $e$, $i$, and $A_{\phi}$ within the resonant phase space for each libration island and a slope of $\alpha=0.9$ for the $H$ distribution (see Section~\ref{s:21} and Figure~\ref{f:21vis} for comparison).}\label{f:visfull21}
\end{figure}

We have presented the detections of resonant objects from the first two of the eight OSSOS observational blocks.  The OSSOS detections of 3:2, 5:2, and 2:1 resonant objects are broadly consistent with the resonance models and population estimates found by CFEPS (G12).  This verification of CFEPS results with an entirely independent dataset inspires additional confidence in the results from the CFEPS/OSSOS survey characterization method.  Our primary results are as follows:
\begin{itemize}
\item Our population estimates are listed in Table \ref{t:pops}.  These values are consistent with CFEPS population estimates within the uncertainties (G12).  We find that given a modified empirical color conversion, the DES population estimates \citep{Adams2014} are also consistent with these results within our 95\% confidence intervals.
\item OSSOS detections of several very low amplitude 3:2 objects require a refinement of the CFEPS plutino model, extending the libration amplitude distribution to lower values.  Lower amplitude librators are produced more efficiently in models appealing to capture during smooth migration of Neptune than in models which fill Kuiper belt phase space (for example during dynamical upheaval of the giant planets) and leave behind resonant populations because the resonances are regions of dynamical stability.  Additional dynamical modeling is required to determine whether our low-amplitude librators provide evidence for a population of migration-captured resonant objects in the 3:2.  Our low-amplitude detections were enabled by placement of the 13AO block $\approx$10 degrees from one of the ortho-Neptune perihelion locations.  Future OSSOS blocks will improve our characterization of this population.  Figure \ref{f:visfull32} displays an estimated visibility map for the 3:2 resonance given the full OSSOS survey.
\item We find no evidence of the H-magnitude distribution transition suggested by \citet{Alexandersen2015} in our population of 3:2 objects. However, we find that if such a transition is present, our small sample of objects would reject a single slope $H$ model only $\sim$80\% of the time.  The increased sample size provided by future OSSOS blocks will place better constraints on the $H$ distribution of the plutinos.
\item The OSSOS 5:2 detections confirm the finding in G12 that this resonance is more populated than expected based on existing models for the dynamical history of the outer Solar System.  After restricting ourselves to the eccentricity range visible in the OSSOS blocks ($e>0.35$), we independently verify that the total population of the 5:2 is at least as large as that of the 2:1 and possibly as large as that of the 3:2.  Given this confirmation, future models of dynamical emplacement of Kuiper belt objects must produce a large population in the 5:2.  The addition of future OSSOS detections will reduce the large uncertainty in our resonant population estimates and allow a more precise measurement of the 3:2 / 5:2 / 2:1 population ratios.  
\item We have confirmed that the inclination distribution of the 2:1 resonance is much colder than those of the 5:2 and 3:2.  This result might indicate that a larger fraction of 2:1 objects were caught in resonance from a dynamically unexcited reservoir.  We speculate that a larger fraction of 2:1 objects may have been caught during migration of Neptune from the objects originally located in the region of the observed cold classical Kuiper belt, although this scenario might not be consistent with the wide range of colors seen for 2:1 objects compared to the cold classicals by \citet{Sheppard2012}.  We will investigate this speculation in future modeling work.
\item Using the combined CFEPS and OSSOS 2:1 detections, we have placed new, more restrictive constraints on both the fraction of the 2:1 resonant objects that are symmetric librators as well as the ratio of leading to trailing asymmetric librators.  Our current limits do not substantially constrain histories of resonance capture during migration, but future OSSOS blocks will be sensitive to both leading and trailing asymmetric librators (see Figure~\ref{f:visfull21} for an estimated 2:1 visibility map for the full survey).  If the populations of leading and trailing librators are significantly different, this difference may be identifiable by the full OSSOS sample.
\end{itemize} 

OSSOS observed two blocks leading Neptune from late 2013-late 2015, and this should slightly more than double the resonant sample once the data is fully analyzed; this will allow an improvement of the current analysis (especially for the 2:1).  The second half of OSSOS will produce orbits by early 2017, and will cumulatively provide $\sim5-6$ times the current 13AE/O block sample; the multiple is $>4$ due to filter upgrades and seeing improvements occurring in a now-vented dome at the CFHT telescope, both of which improve magnitude depth.

\acknowledgements{\noindent K. Volk and R. Murray-Clay are supported by NASA Solar System Workings grant number NNX15AH59G.  This research was also supported by funding from the National Research Council of Canada and the National Science and Engineering Research Council of Canada. Based on observations obtained with MegaPrime/MegaCam, a joint project of the Canada--France--Hawaii Telescope (CFHT) and CEA/DAPNIA, at CFHT which is operated by the National Research Council (NRC) of Canada, the Institute National des Sciences de l'Universe of the Centre National de la Recherche Scientifique (CNRS) of France, and the University of Hawaii. The authors recognize and acknowledge the sacred nature of Maunakea, and appreciate the opportunity to use data observed from the mountain. This work is based in part on data produced and hosted at the Canadian Astronomy Data Centre. We acknowledge useful discussions with Rebekah Dawson about the use of statistical tests in this work.}

\appendix

\section{Statistical Tests}\label{a:stats}

Throughout this paper, we describe the $H$ distribution and dynamical properties of the underlying resonant population using simple parametrized models.  Given the set of objects observed by OSSOS, we would like to determine which values of the models' parameters are most probable, identify the range of parameter values that reasonably match the data, and verify that our simple models can fit the data sufficiently well that more complicated models are not required.  

To achieve the first two goals, we would in principle like to calculate, in multiple dimensions, the relative likelihoods of observing our detected objects given each set of model parameters.  Given a uniform (uninformative) prior, these relative likelihoods are equivalent to the Bayesian posterior distribution or, in other words, the relative likelihoods of each set of model parameters.  The sufficiency of our models could then be assessed by using the most probable model parameters and comparing the probability computed for the observed data set to the distribution of probabilities produced by synthetic data sets.  This procedure is described in Section \ref{sec-maxlike}.  In practice, this full calculation is difficult because significant computational resources are required to evaluate the observational biases in our data.  

Fortunately, our inferred distributions for the inclination and libration amplitude do not depend substantially on the inferred distributions of other parameters (see Section \ref{ss:32iaphi}).  We can therefore employ a maximum likelihood calculation for the inclination distribution and libration amplitude distributions using 1-dimensional models, fixing all other distributions to a set of acceptable parameter values.

The inferred distance of detection, absolute magnitude, and eccentricity distributions, however, are correlated and must be treated together.  Even this three dimensional space is very computationally expensive to probe with high resolution using a relative likelihood calculation (Section \ref{sec-maxlike}), so we compromise by combining an Anderson-Darling rejectability statistic and a $\chi$-squared calculation intended to assess goodness-of-fit, described in Sections \ref{sec-AD} and \ref{sec-CS}, respectively.

\subsection{Maximum Likelihood}\label{sec-maxlike}

From Bayes' theorem, the probability of a model $A$ given our data set $D$ is $P(A|D) \propto P(D|A)P(A)$.  Our model, $A$, consists of a set of parameters that characterize the distributions of eccentricity ($e$), inclination ($i$), libration amplitude ($\phi$), absolute magnitude ($H$), and heliocentric distance at discovery ($d$) for the KBO population in a specified resonance.  In all cases, the logarithmic scale of our model parameters is known.  We therefore assume uniform (uninformative) priors on our parameterized models $P(A)$, so that
\begin{equation}
P(A|D) \propto P(D|A) \;\;.
\end{equation}
To calculate the relative likelihoods of each model (each set of parameterized distributions) given our set of observed objects $D$,  we would therefore like to know the probability of observing $D$ given each possible set of model parameters.  

The nature of the observational biases that must be applied to our models means that we cannot analytically calculate $P(D|A)$.  Instead, we must estimate this probability numerically.  To do so, we use the OSSOS survey simulator.  The survey simulator produces a set of synthetic detections given a model of the underlying population.  If the survey simulator could be run an infinite number of times, it would (when normalized) translate each model (or set of parameter values) into a probability distribution for the properties of a detected object, $P_1(e, i, \phi, H, d | A)$, given the model and a specified total number of detections; the total number of objects in the model itself is a free parameter in our parameterized models, and it is allowed to vary so that the model produces a total number of detections equal to the actual number of detections. The probability of observing $N$ detected objects with a given set of $(e, i, \phi, H, d )$ is then
\begin{equation}\label{eq:pda}
P(D|A) \propto \prod_{j=1}^{N}P_1(e_j, i_j, \phi_j, H_j, d_j | A)
\end{equation}
The most probable model is the one that maximizes $P(D|A)$.   

In practice, we can only run the survey simulator a finite number of times per set of model parameters, so our calculation of $P_1$ must be binned in $e$, $i$, $\phi$, $H$, and $d$.   Because we are calculating $P_1(D|A)$ numerically from synthetic detections, there is an uncertainty due to Poisson noise in the number of detections expected by the model.  In bin $k$, we estimate this uncertainty to be
\begin{equation}
dP_1(x_k | A) = \frac{\sqrt{n_{k}}}{n_{k}}P_1(x_k | A)
\end{equation}
where $n_{k}$ is the number of synthetic detections in bin $k$ and we have used $x_k$ as shorthand for the values of $(e,i,\phi,H,d)$ in each bin. We then propagate this uncertainty through the calculation of $P(D|A)$  (equation~\ref{eq:pda}) by adding them in quadrature.  Two models can only be distinguished if their relative probabilities $P(A|D)$ differ by more than the errors in their $P(A|D)$.

In principle, we would like to---for each set of parameter values---produce sufficiently many simulated detections that we could map this five-dimensional probability distribution, allowing us to directly compute the relative probabilities of the observed data for a large sets of parameter values spanning their full range and allowing for correlations in the parameters.  In practice, running the survey simulator a sufficient number of times to do this is prohibitively computationally expensive given the size of the currently allowed parameter space (see \citealt{mikephd} for another discussion of this); when OSSOS is complete, the allowed parameter space will be smaller and there will likely be enough real detections to make the expenditure of computational resources worthwhile. Due to the computational expense, in this paper we only apply this approach in a few cases. For our parameterized inclination distribution, we can do this calculation in one dimension (inclination) to find the most probable inclination width; we can limit ourselves to one dimension because the observability of the inclination distribution is fairy independent of the other orbital element distributions and the $H$ distribution (see Figure~\ref{f:nocouple}).  In that case, we fix the eccentricity, $h$, and libration amplitude distributions and run a suite of survey simulations over a wide range of inclination widths to find which width, $\sigma_i$, maximizes the probability $P(\sigma_i |D) \propto  \prod_{j=1}^{N} P_1(i_j | \sigma_i)$.   A similar calculation is done for the parameterized libration amplitude distribution.  These one-dimensional probability calculations only require $\sim10^4-10^5$ synthetic detections per model iteration to adequately sample the binned probability distribution with fractional uncertainties $<10^{-3}$ in $P_1$ for almost all bins that contain a real detection.  

To assess model sufficiency for the inclination distribution (and analogously for the libration amplitude distribution), for each $\sigma_i$, we produce a large set of synthetic data sets of size $N$ drawn from equation~\ref{eq:i}.  We calculate $P(D|A)$ for each synthetic data set using equation~\ref{eq:pda}.  We then determine whether $P(D|A)$ calculated for the observed data is consistent with the distribution of $P(D|A)$ generated by the synthetic data.  Consistency is defined as lying in the middle 95\% of the cumulative distribution (removing the bottom 2.5\% and top 2.5\%).  We find that the range of values for $\sigma_i$ accepted by this procedure is similar to the range of values not rejected by the Anderson-Darling test, described in Appendix \ref{sec-AD}.

Because of the coupled observational biases, calculating relative probabilities for our different $H$ distribution/eccentricity models requires us to bin the probability distribution in three dimensions: eccentricity, distance at discovery, and absolute magnitude.  For reasonable bin sizes, determining the probability distribution to fractional uncertainties $<10^{-2}$ in most bins, and $<10^{-1}$ in all bins requires of order $10^7$ synthetic detections per model iteration.  For this reason, we only calculate $P(D|A)$ for a few of our favored models in order to compare their relative probabilities.

\subsection{Anderson-Darling Test}\label{sec-AD}

We use the Anderson-Darling test statistic (AD statistic) outlined by NIST\footnote{\url{http://www.itl.nist.gov/div898/handbook/eda/section3/eda35e.htm}}.  The AD statistic is defined as
\begin{equation}
A^2 = -N - S
\end{equation}
where there are $N$ ordered data points, $Y_j$, and $S$ is given by:
\begin{equation}
S = \sum_{j=1}^{N}{\frac{(2j-1)}{N}\left[ lnF(Y_j) + ln(1-F(Y_{N+1-j})) \right]}
\end{equation}
where $F(Y_j)$ is the cumulative distribution function of the model being tested.  Larger values for $A^2$ correspond to cumulative distributions that are more different.  The significance of the value of $A^2$ is determined by calculating the expected distribution of $A^2$ by repeatedly drawing samples of $N$ random points from the model distribution and computing their AD statistic.  A model is rejected at P\% confidence if the observed value of the AD statistic is larger than P\% of the model's subsample AD statistics.

The above calculation is for a one-dimensional distribution.  To extend to testing multiple distributions at once, we linearly add the AD statistic for each one-dimensional distribution \citep{Parker2015,Alexandersen2015}.  The significance of the summed AD statistic is then determined just as above.  

\subsection{Chi-square Statistic}\label{sec-CS}

One measure of the goodness-of-fit for any particular parameterized model is the chi-square statistic:
\begin{equation}
\chi ^2 = \sum_{j=1}^{n}(O_j - E_j)^2/E_j
\end{equation}
where $O_j$ is the observed frequency in bin $j$ and $E_j$ is the expected frequency in bin $j$ given the model \citep{Press1992}.  The exact value of $\chi^2$ will depend on the choice of bins for the data, but if the binning is the same across multiple models, a smaller value of $\chi^2$ indicates a better fit to the observations.   As with the AD statistic, a simplistic way to extend this to multiple dimensions is to linearly add the one-dimensional values of $\chi^2$.  We do not have enough observational data points to calculate a meaningful $\chi^2$ by binning in multiple dimensions simultaneously.

\section{Orbit Fitting and Uncertainties}\label{a:orbits}

The determination of the best-fit orbit and classification status of each OSSOS detection follows the procedure outlined in \citet{Gladman2008}.  A barycentric best-fit orbit is determined based on all available OSSOS astrometric observations using the \citet{Bernstein2000} algorithm.  This is the best-fit orbit listed in Table~\ref{t:detections}; the listed uncertainties are taken from the diagonal elements of the best-fit orbit's covariance matrix.  To determine the classification status of an object, we integrate three clones of the observed object to determine the dynamical behavior: the best-fit orbit and then orbits with the maximum and minimum deviations in semimajor axis that are still consistent with the observations.  To generate the minimum- and maximum-$a$ clones, we perform a search for acceptable orbits in \citet{Bernstein2000}'s ($\alpha,\beta,\gamma$) coordinate system, with a maximum variation in each coordinate of $3\sigma$ as determined from the diagonal elements of the ($\alpha,\beta,\gamma$) covariance matrix.  An orbit fit is deemed consistent with the observations if the worst residual between its predictions and the observed astrometric positions is not larger than 1.5 times the best-fit orbit's worst residual and if the rms residual is not more than 1.5 times the best-fit orbit's rms residual.  The \citet{Bernstein2000} orbit fitting code does provide a semimajor axis uncertainty ($\sigma_a$), but \citet{Gladman2008} chose to use the residuals to determine the minimum- and maximum-$a$ orbits because doing so provides a better estimate of the true uncertainty when the observational arc is short or when there might be systematic errors in the astrometry.  In these cases, it is not unusual for the maximum- and minimum-$a$ orbits to differ from the best-fit value by significantly more than $3\sigma_a$.  This is because the nominal 1-$\sigma_a$ uncertainty for the best-fit orbit is calculated from the diagonals of the barycentric orbital elements covariance matrix, which is produced by assuming an linearized conversion from the ($\alpha,\beta,\gamma$) coordinate system to orbital elements; in cases where the orbital uncertainties are still large, this conversion can become inaccurate \citep{Bernstein2000}, leading to the diagonal elements underestimating the uncertainty in $a$.  The accuracy of the OSSOS astrometry means that for most of the objects listed in Table~\ref{t:detections}, the minimum- and maximum-$a$ orbits converge to within $3\sigma_a$ of the best-fit $a$ even with a relatively short observational arc; however, we still use the \citet{Gladman2008} procedure for assessing the acceptable range in $a$.

The best-fit, minimum-$a$, and maximum-$a$ orbits are integrated forward in time for 10~Myr using SWIFT \citep{Levison1994} under the influence of the Sun and the four giant planets.  We check for libration of any resonance angle (defined by equation~\ref{eq:full_phi}) for all $p:q$ resonances with $p\le30$ and  $|p-q|<30$ within 2\% of the best-fit $a$.  An object is securely resonant if all three clones librate for more than half of the 10 Myr integration.  Insecure resonant objects (indicated by `I' in Table~\ref{t:detections}) have a best-fit orbit that is resonant but at least one other clone that does meet this criterion. Objects listed as `IH' in Table~\ref{t:detections} had to be classified by hand as resonant because the libration behavior of the clones was too messy to be correctly identified as libration by the automated code.

\begin{figure}[htbp]
   \centering
   \includegraphics[width=6.2in]{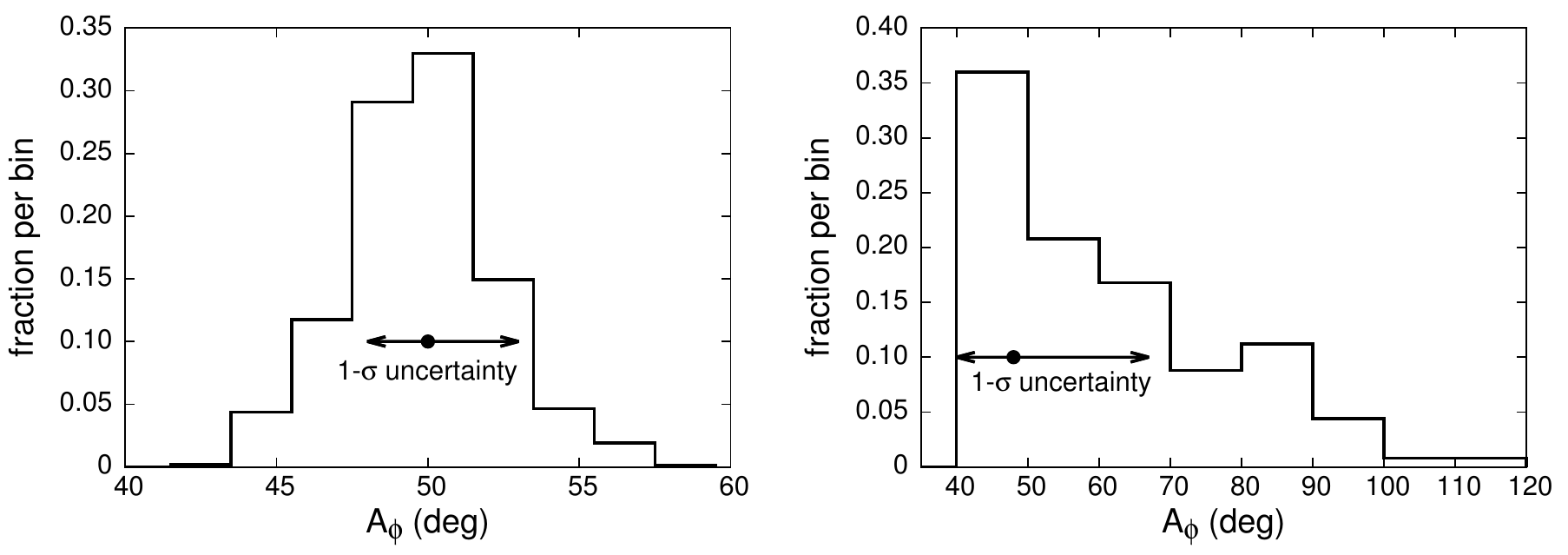}
   \caption{Distributions of $A_{\phi}$ for two plutinos.  The left panel shows OSSOS plutino o3o15, where the $A_{\phi}$ distribution is nearly symmetric around the best-fit orbit's $A_{\phi}$. The right panel shows o3o20PD, which has a very asymmetric $A_{\phi}$ distribution.  In both panels the best-fit orbit's $A_{\phi}$ is indicated by a black circle and the arrows show what we are calling the $1\sigma$ uncertainties.}
   \label{af:aphi-dist}
\end{figure}

To better assess the likelihood that an insecure object is resonant and to determine the uncertainty in the libration amplitude for securely resonant objects, we integrated a distribution of clones for each resonant detection.  We generated 250 orbits by using the full ($\alpha,\beta,\gamma$) covariance matrix to produce Gaussian deviations from the best-fit orbit's ($\alpha,\beta,\gamma$).  We integrated all 250 clones for 10 Myr and measured the libration amplitude, $A_{\phi}~=~(~\phi_{max}~-~\phi_{min})/2$, over ten 1 Myr windows and averaged the ten $A_{\phi}$ values.  The resulting distribution in $A_{\phi}$ is often not symmetric around the best-fit orbit's $A_{\phi}$.  In cases where the distribution is roughly symmetric about the best-fit $A_{\phi}$, we use the cumulative distribution in $A_{\phi}$ to find the $1\sigma$ values that bracket the central 67\% of the distribution.  In cases where it is highly asymmetric about the best-fit (see Figure~\ref{af:aphi-dist}), we take the $1\sigma$ uncertainty to be the range from the minimum $A_{\phi}$ to the point in the cumulative distribution where 67\% of the clones have smaller $A_{\phi}$ (or the reverse if the asymmetric peak in the $A_{\phi}$ distribution occurs at large $A_{\phi}$).  In some cases where the libration behavior is not well-behaved and the test particles slip in and out of libration, the time histories are examined manually to determine the percentage of the clones that are resonant for at least half of the 10 Myr simulation; in these instances no value of $A_{\phi}$ is given in Table~\ref{t:detections}, and just the percentage of resonant clones is listed.

\section{Survey Simulator Details: Modeling the Resonances}\label{a:resmodeling}

This section gives a more detailed accounting of the modeling of each resonant population within the OSSOS survey simulator.  In the plutino section below, we also outline some validation testing we have performed to ensure that the simplified, parameterized models of the resonances are adequate representations of the resonant populations given the current data.  In all cases below, we select orbital elements for the epoch JD 2453157.5, at which time Neptune's mean longitude is $\lambda_N = 5.489$ (the survey simulator propagates the orbits to the appropriate epoch for each OSSOS observation).

\subsection{Modeling the 3:2 population}\label{as:32}

The first step in generating plutinos within the survey simulator is to choose whether the plutino is also in the Kozai resonance.  For non-Kozai plutinos, the procedure for choosing an orbit is as follows:
\begin{itemize}
\item $a$ is randomly chosen uniformly in the range $39.45\pm0.2$~AU
\item $e$ is chosen randomly from equation~\ref{eq:e}
\item $i$ is chosen randomly from equation~\ref{eq:i}
\item the object's $A_{\phi}$ is chosen randomly from the specified distribution (see Section~\ref{ss:32iaphi})
\item $\phi$ is then given by $\phi = 180^{\circ} + A_{\phi} \sin(2\pi t)$ where $t$ is a random number distributed uniformly from $0-1$
\item $M$ is randomly chosen uniformly in the range $0-2\pi$
\item $\Omega$ is randomly chosen uniformly in the range $0-2\pi$
\item $\omega$ is fully constrained by the above choices and is given by $\omega = \frac{1}{2}\phi-\frac{3}{2}M - \Omega + \lambda_N$
\end{itemize}
The object's absolute magnitude is then chosen randomly from equation~\ref{eq:spl}, fully specifying the object's position and brightness.  The procedure for selecting $a$ and $e$ is simplified slightly compared to that in G12 and \citet{Alexandersen2015}, who included the shape of the resonance's $a-e$ phase space in this selection by rejecting (and re-selecting) $a$ and $e$ in instances where the selection falls outside the resonance boundaries; we find that this complication is unnecessary because the exact value of $a$ does not change the observability of an object in the survey simulator (in fact, the results of our modeling would not change if we completely eliminated the variations in $a$ and just assigned every plutino $a=39.45$ AU).

This procedure is simplified compared to real resonant dynamics, so we have performed a few basic tests to ensure that it provides an adequate representation of the plutino population.  The two major simplifications made above are (1) the independent selection of $e$, $i$, and $A_{\phi}$ and (2) using a simple harmonic oscillator to represent the time evolution of $\phi$ (equation~\ref{eq:phitime}) while ignoring the small changes in osculating orbital elements (particularly $e$ and $i$) that occur over a resonant cycle.

\begin{figure}[htbp]  \centering
   \includegraphics[width=6.0in]{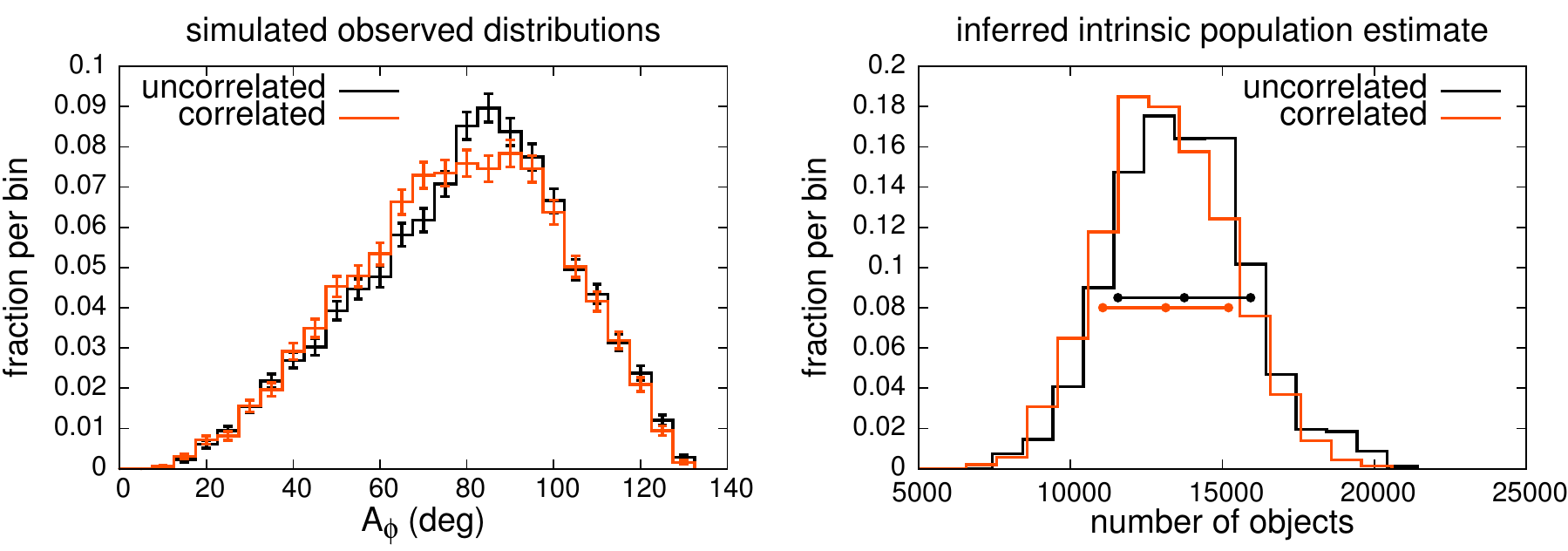}
\caption{Left: Distribution of simulated detections for populations with and without a correlation between libration amplitude and eccentricity. Right: Distribution of total population estimates for the two cases. The two horizontal lines indicate the average and 1-$\sigma$ population estimates.}
\label{f:correlation_example}
\end{figure}

To test the implications of the first simplifying assumption, we constructed various plutino populations that have identical 1-dimensional distributions of $e$, $i$, and $A_{\phi}$, but within the generating procedure we imposed different relationships between these distributions to see how the relationship affects the distribution of simulated detections and the total population estimates generated by the survey simulator.  An example: depending on how an object becomes trapped in resonance, there can be a relationship between its eccentricity and its libration amplitude \citep[e.g.,][]{Chiang2002}.  A correlation between libration amplitude and eccentricity in the real population of plutinos could introduce observational biases that would not be properly accounted for in a survey simulator model that treats those two distributions independently.  The result could be a model where the real and simulated detections match in terms of the individual, 1-dimensional distributions but has an inaccurate total population estimate because the correlation was not modeled.  Figure~\ref{f:correlation_example} shows two plutino models with identical intrinsic $e$ and $A_{\phi}$ distributions: one model has a population in which low libration amplitude simulated objects have higher eccentricities than large libration amplitude objects and the other model has no correlation.  In this case there are small differences between the simulated detected $A_{\phi}$ distributions for the two models and a small difference in the total population estimate; note that the pattern of small differences here will depend on the sky coverage of a given survey.  However the differences are much smaller than the model uncertainties given a hypothetical observed sample size of 100.  A test of correlations between $i$ and $A_{\phi}$ produces similar results.  Given the current sample size of 21 OSSOS detections and 24 CFEPS detections for the plutinos (and smaller sample sizes for all the other resonances), the simplified, independent treatment of the $e$, $i$, and $A_{\phi}$ distributions is adequate.

\begin{figure}[htbp]
   \centering
   \includegraphics[width=6.2in]{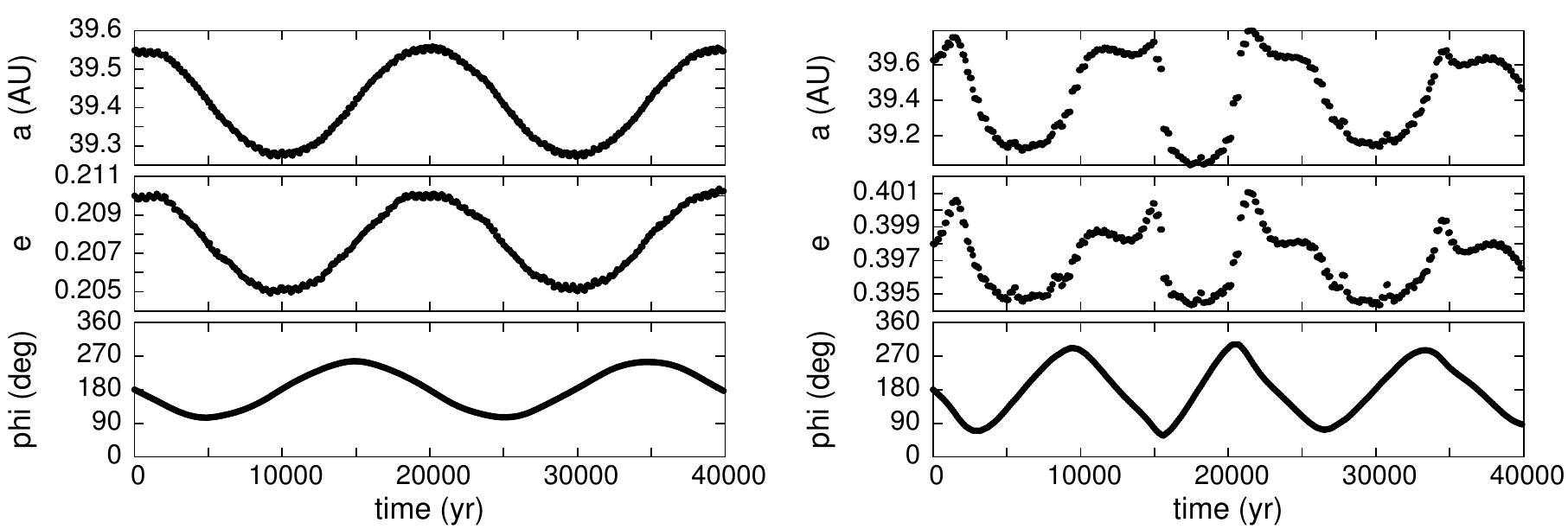}
\caption{Test particles in the 3:2 resonance from a numerical simulation that show sinusoidal (left) and non-sinusoidal (right) variations in $a$,$e$, and $\phi$.}
\label{f:res_cycle}
\end{figure}

To examine our second simplifying assumption, we tested whether more accurately generating objects with positions and velocities consistent with real resonant behavior can affect the results of model testing.  Within the survey simulator, resonant objects are assigned values of ($a$,$e$,$i$,$\Omega$,$\omega$,$\phi$) randomly from within the desired distributions, and then those values are interpreted as instantaneous osculating elements that can be translated into a sky position.  In reality, however, an object's osculating $a$,$e$,$i$ vary depending on where the object is in the resonant cycle, as shown in Fig~\ref{f:res_cycle}.  This means there is a relationship between the exact current osculating $a$,$e$,$i$ and the value of $\phi$ (which determines $\Omega$, $\omega$, and $M$) that is not correctly modeled when these parameters are chosen independently.  The time weighting of $\phi$ for a real population can also differ from the simple harmonic oscillator model we use in the survey simulator.  Figure~\ref{f:res_cycle} shows the variation of $a$, $e$, and $\phi$ for two different plutinos from a numerical integration; the left panel shows a plutino with moderate $A_{\phi}$ and sinusoidal variations while the right panel shows a larger $A_{\phi}$ plutino with non-sinusoidal variations.  The non-sinusoidal $\phi$ variations in the real plutinos means that the survey simulator with its simplified sinusoidal time-weighting is not correctly accounting for the amount of time a real plutino spends at specific values of $\phi$ (which determines where on the sky relative to Neptune the object's perihelion most often occurs). To see if more accurate time-weighting of $\phi$ and the inclusion of the associated variations in $a$ and $e$ would significantly alter our conclusions about the plutino population, we performed short numerical integrations of $i\sim0$ plutinos whose time histories can be used to generate plutinos within the survey simulator.  We then took our nominal plutino $H$, $e$, and $A_{\phi}$ model from Section~\ref{s:32} and generated objects in the survey simulator either by our standard procedure or by selecting objects' orbital elements from the numerical simulations.  In this planar test case, we find that sampling from the numerical integrations does not meaningfully change the distribution or number of simulated observed objects.  There are also no disallowed combinations of $e$, $i$, and $A_{\phi}$ in the planar case (i.e. all values of $e$ exist in resonant phase space for all values of $A_{\phi}$), so our independent treatment of these distributions is not artificially generating plutinos that would actually be in a non-resonant portion of the phase space.

The procedure for Kozai plutinos is different because the Kozai resonance presents an additional constraint on the orbit due to the libration of $\omega$ around either $90^{\circ}$ or $270^{\circ}$.  In the idealized three body problem, where we just consider the evolution of a test particle under the influence of the Sun and Neptune, a plutino librating in the Kozai resonance will follow a closed path in $e$-$\omega$ space (see for example Figure 4 in \citealt{Wan2007}).  This trajectory represents a path in $e$-$\omega$ space with a constant value of the resonant Hamiltonian (see for example the expression for the approximate Hamiltonian given in \citealt{Wan2007}).  Sets of Hamiltonian level curves (and thus libration trajectories) can be grouped according to the value of the z component (normal to the reference plane) of the orbit's angular momentum, $L_z \propto \sqrt{1-e^2}\cos i$.  $L_z$ is a preserved quantity, meaning that a test particle's eccentricity and inclination evolution are coupled as the test particle traces out its trajectory in $e$-$\omega$ space.  A plutino's Kozai libration behavior could be specified by selecting $L_z$, calculating all of the associated Kozai resonant Hamiltonian level curves for that $L_z$, selecting one of these level curves, selecting a ($e$,$\omega$) pair from that level curve trajectory, and then calculating the appropriate $i$ based on $e$ and $L_z$.   One could then select a 3:2 libration amplitude, set $\phi$ as above in the non-Kozai case, select a random $M$, and calculate $\Omega =\frac{1}{2}\phi-\frac{3}{2}M -\omega + \lambda_N$ to fully specify the orbit and position (this is the procedure used in \citealt{Lawler2013}).  

\begin{figure}[htbp]
   \centering
   \includegraphics[width=6.2in]{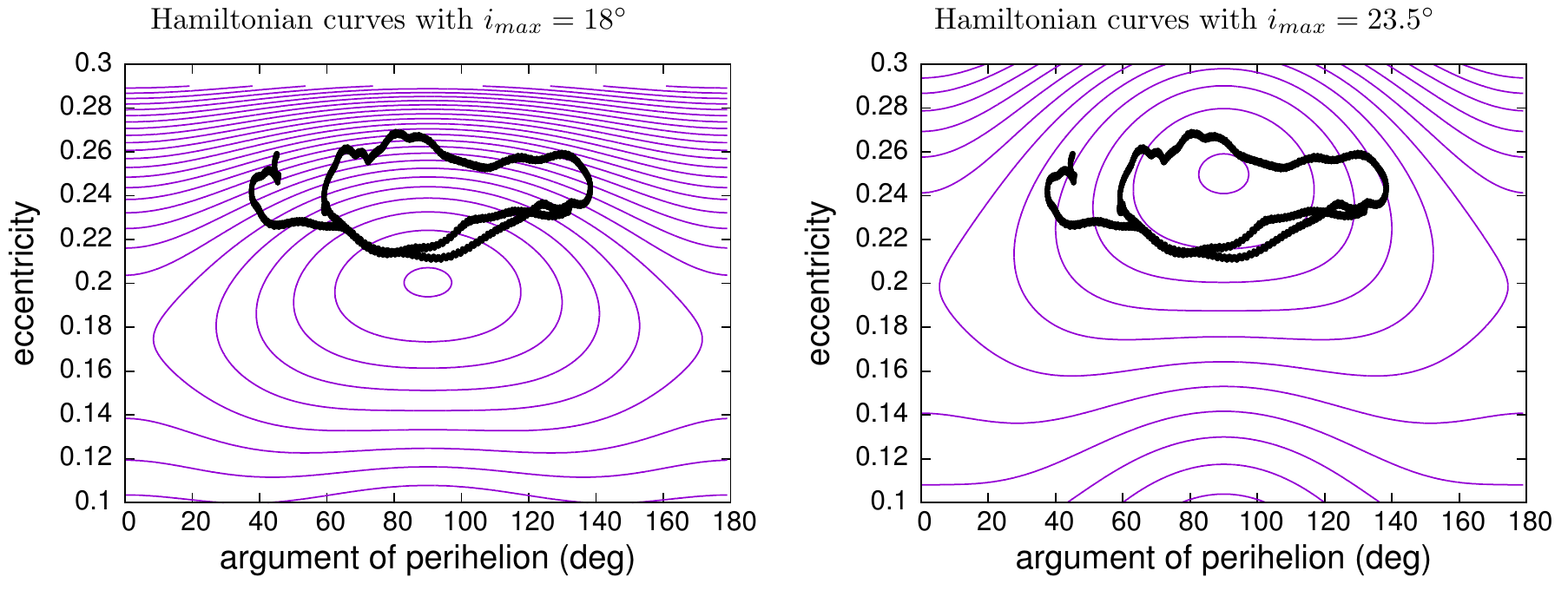}
   \caption{Evolution of plutino o3o08 (black dots) from a numerical simulation that includes the Sun and the four outer planets plotted over approximate Hamiltonian level curves for $i_{max} = 18^{\circ}$ (left panel) and $i_{max} = 23.5^{\circ}$ (right panel).  The object has $i_{max} = 18^{\circ}$, but the poor match in the left panel indicates that the expression for the approximate Hamiltonian in the three body problem as given by \citet{Wan2007} is not a good match to the dynamics in the full integrations.}
   \label{af:kz-example}
\end{figure}

The above procedure would be the best way to generate Kozai plutinos in the three body problem, but the real Kozai plutinos do not follow such a nicely defined set of trajectories.  In Figure~\ref{af:kz-example} we show the evolution of one of the OSSOS Kozai plutinos in the 10 Myr classification integration superposed over Hamiltonian level curves calculated using the approximate Hamiltonian from \citet{Wan2007} for two different values of $L_z$; the values of $L_z$ are parameterized by $i_{max}$, the maximum inclination allowed by conservation of $L_z$ at $e=0$.  The observed plutino in Figure~\ref{af:kz-example} has $i_{max} = 18^{\circ}$, but the figure shows that its evolution in $e-\omega$ space is not well described by any of the $i_{max} = 18^{\circ}$ Hamiltonian level curves.  From visual inspection of the evolution of the five OSSOS Kozai plutinos, we find that all of the Kozai plutinos have an average eccentricity over their Kozai cycles of $e\sim0.25$ despite having $i_{max}$ ranging from $17-24^{\circ}$.  The three body Hamiltonian predicts that the average $e$ should decrease with decreasing $i_{max}$, but the real Kozai plutinos do not show this and all have average $e$ that approximately matches the Hamiltonian level curves for $i_{max}=23.5^{\circ}$ (see the right panel of Figure~\ref{af:kz-example}).  The exact path a Kozai plutino follows in $e-\omega$ space does change with changing $i_{max}$.  The amplitude of the eccentricity variations in the real Kozai plutinos seems to decrease with decreasing $i_{max}$, which can be seen in the right panel of Figure~\ref{af:kz-example} where the real $i_{max}=18^{\circ}$ object's path is flattened relative to the $i_{max}=23.5^{\circ}$ Hamiltonian curves; a real object with $i_{max}=23.5^{\circ}$ would more closely match the eccentricity amplitude predicted by the level curves.

Because all of the OSSOS Kozai plutinos appear to share an average eccentricity with the $i_{max}=23.5^{\circ}$ Hamiltonian curves, we use this one set of level curves to choose $e$ and $\omega$ for all of the Kozai plutinos in the survey simulator.  This is the same approach used in G12 because all of their Kozai plutinos were well described by those level curves and all had $i_{max}$ near $23.5^{\circ}$.  Some of the OSSOS Kozai plutinos have significantly lower current orbital inclinations and $i_{max}$ than the G12 Kozai plutinos, so we have altered the way the inclinations of Kozai plutinos are calculated within the survey simulator.  After $e$ and $\omega$ are chosen from the $i_{max}=23.5^{\circ}$ level curves, we then choose a new value of $i_{max}$ uniformly in the range $17-24^{\circ}$ and then set the generated Kozai plutino's inclination such that $\cos i = (\cos i_{max})/(\sqrt{1-e^2})$.  Thus the full procedure we use for generating a Kozai plutino is as follows:
\begin{itemize}
\item $a$ is randomly chosen uniformly in the range $39.45\pm0.2$~AU
\item a Hamiltonian level curve is randomly chosen from the Kozai resonant $i_{max}=23.5^{\circ}$ level curves
\item a position on that level curve is randomly chosen (with the simplified assumption that equal time is spent on all parts of the level curve), setting $e$ and $\omega$
\item $i_{max}$ is chosen randomly from the uniform range $17-24^{\circ}$
\item $i$ is then calculated from $\cos i = (\cos i_{max})/(\sqrt{1-e^2})$
\item the object's $A_{\phi}$ is chosen randomly from the specified distribution (see Section~\ref{ss:32iaphi})
\item $\phi$ is given by $\phi = 180^{\circ} + A_{\phi} \sin(2\pi t)$ where $t$ is a random number distributed uniformly from $0-1$
\item $M$ is randomly chosen uniformly in the range $0-2\pi$
\item $\Omega$ is fully constrained by the above choices and is given by $\Omega =\frac{1}{2}\phi-\frac{3}{2}M -\omega + \lambda_N$
\end{itemize}
The object's absolute magnitude is then chosen randomly from equation~\ref{eq:spl}, fully specifying the object's position and brightness.  This procedure is not entirely self-consistent in the use of analytical models of the Kozai population, but as shown in Figure~\ref{af:kz-example}, the real evolution of the Kozai plutinos matches the Hamiltonian level curves in a qualitative rather than quantitative sense.  This procedure yields synthetic detected Kozai plutinos that are an acceptable match to the real detections.  In future work the procedure will be re-evaluated to see if a more accurate representation of Kozai phase space is required.

\subsection{Modeling the 5:2 population}\label{as:52}

The procedure for choosing an orbit for a 5:2 object is as follows:
\begin{itemize}
\item $a$ is randomly chosen uniformly in the range $55.4\pm0.2$~AU
\item $e$ is chosen randomly from equation~\ref{eq:e}
\item $i$ is chosen randomly from equation~\ref{eq:i}
\item the object's $A_{\phi}$ is chosen randomly from the specified distribution (see Section~\ref{s:52})
\item $\phi$ is then given by $\phi = 180^{\circ} + A_{\phi} \sin(2\pi t)$ where $t$ is a random number distributed uniformly from $0-1$
\item $M$ is randomly chosen uniformly in the range $0-2\pi$
\item $\Omega$ is randomly chosen uniformly in the range $0-2\pi$
\item $\omega$ is fully constrained by the above choices and is given by $\omega = \frac{1}{2}\phi-\frac{5}{2}M - \Omega + \lambda_N$
\end{itemize}
The object's absolute magnitude is then chosen randomly from equation~\ref{eq:spl}, fully specifying the object's position and brightness.

\subsection{Modeling the 2:1 population}\label{as:21}

To generate an orbit for a 2:1 resonant object, we first decide if it is a symmetric or asymmetric librator.  If it is a symmetric librator, we then choose the orbit as follows
\begin{itemize}
\item $a$ is randomly chosen uniformly in the range $47.8\pm0.2$~AU
\item $e$ is randomly chosen uniformly from $0.05-0.35$
\item $i$ is chosen randomly from equation~\ref{eq:i}
\item $A_{\phi}$ is randomly chosen uniformly from $125-165^{\circ}$
\item $\phi$ is then given by $\phi = 180^{\circ} + A_{\phi} \sin(2\pi t)$ where $t$ is a random number distributed uniformly from $0-1$
\item $M$ is randomly chosen uniformly in the range $0-2\pi$
\item $\Omega$ is randomly chosen uniformly in the range $0-2\pi$
\item $\omega$ is fully constrained by the above choices and is given by $\omega = \phi- M - \Omega + \lambda_N$
\end{itemize}
For asymmetric librators, we choose the orbit as follows:
\begin{itemize}
\item $a$ is randomly chosen uniformly in the range $47.8\pm0.2$~AU
\item $i$ is chosen randomly from equation~\ref{eq:i}
\item $e$ is randomly chosen uniformly from $0.1-0.4$
\item the center of libration, $\phi_c$, for that eccentricity is calculated from Figure 4 in \citet{Nesvorny2001}
\item the maximum $A_{\phi}$, for that eccentricity is also calculated from Figure 4 in \citet{Nesvorny2001}
\item $A_{\phi}$ is randomly chosen uniformly from $0-A_{\phi, max}$
\item the object is chosen to be in either the leading or trailing asymmetric island and $\phi_c$ is adjusted appropriately
\item $\phi$ is then given by $\phi = \phi_c + A_{\phi} \sin(2\pi t)$ where $t$ is a random number distributed uniformly from $0-1$
\item $M$ is randomly chosen uniformly in the range $0-2\pi$
\item $\Omega$ is randomly chosen uniformly in the range $0-2\pi$
\item $\omega$ is fully constrained by the above choices and is given by $\omega = \phi- M - \Omega + \lambda_N$
\end{itemize}
The object's absolute magnitude is then chosen randomly from equation~\ref{eq:spl}, fully specifying the object's position and brightness.  The above ranges in $e$ and $A_{\phi}$ for the symmetric and asymmetric librators are chosen based on numerical modeling and stability analysis of the 2:1 resonance by \citet{Nesvorny2001} and \citet{Tiscareno2009}.

\section{Colors for comparison to the DES}\label{a:colors}

In order to compare population estimates for the resonant populations from the DES \citep{Adams2014}, CFEPS (G12), and OSSOS \citep{Bannister2015}, we have to know how to compare the $H$ magnitudes of the resonant objects in the surveys' different filters.  $H$ magnitudes for the DES objects are given in the $VR$ band in \citet{Adams2014}, whereas the $H$ magnitudes of the CFEPS detections are in $g$ band \citep{Petit2011} and the OSSOS detections are in $r$ band.  As discussed in Section~\ref{ss:32Hd}, the average color for comparing CFEPS and OSSOS resonant objects is $g-r=0.5$ based on CFEPS objects that were observed in both $r$ and $g$ band \citep{Petit2011}. Fortunately, some objects seen in the DES were serendipitously present in the CFEPS fields, allowing a direct estimation of the average color is between the g and VR photometric systems (light curves, while present, should average out).  We compared the list of CFEPS resonant detections and DES resonant detections and found all instances of overlapping detections where both $H_{VR}$ and $H_g$ are measured.  These objects are listed in Table~\ref{at:colors}.  We find that these resonant objects have colors in the range $g-VR=0.04-0.86$ with an average color of $g-VR=0.4$.  This differs from the assumption of $g-VR=0.1$ in \citet{Adams2014}; although $g-VR = 0.1$ falls inside the lower extremity of the range, typical colors are much larger.  As such, a comparison of population estimates requires a larger shift between the surveys. 

\begin{deluxetable}{c l l l l l}
\tabletypesize{\footnotesize}
\tablecolumns{6}
\tablewidth{0pt}
\tablecaption{Resonant Object Colors}
\tablehead{\colhead{Res} & \colhead{CFEPS designation} & \colhead{MPC designation} & \colhead{$H_g$} & \colhead{$H_{VR}$} & \colhead{$H_g$ -  $H_{VR}$} \\}
\startdata
5:2	&L4j06PD	&2002 GP32	&	7.03			&6.99	&0.04\\
5:2	&L4h02PD&	2004 EG96	&	8.36		&	8.01	&0.35\\
7:4	&K02O03	&2000 OP67		&8.13	&7.27	&0.86\\
2:1	&L4v06	&2004 VK78		&8.5		&	8.16	&0.34\\
7:3	&L5c19PD&	2002 CZ248	&	8.5	&		8.27	&0.23\\
3:2	&L5i06PD	& 306792 (2001 KQ77)	&7.48&			7.2	&0.28\\
3:2	&L4v13	&2002 VV130		&7.6		&	7.51	&0.09\\
12:5	&L5c12	& 119878 (2002 CY224)	&	6.69		&	6.1	&0.59\\
3:1	&L4v08	&307463 (2004 VU130)	&	6.95	&		6.1&	0.85\\
\enddata
\tablecomments{$H$ magnitudes for the resonant objects detected by both CFEPS and DES. $H_g$ measurements are taken from \citet{Petit2011} and $H_{VR}$ are taken from \citet{Adams2014}.}
\label{at:colors}
\end{deluxetable}


\clearpage


\begin{thebibliography}{54}
\expandafter\ifx\csname natexlab\endcsname\relax\def\natexlab#1{#1}\fi

\bibitem[{{Adams} {et~al.}(2014){Adams}, {Gulbis}, {Elliot}, {Benecchi},
  {Buie}, {Trilling}, \& {Wasserman}}]{Adams2014}
{Adams}, E.~R., {Gulbis}, A.~A.~S., {Elliot}, J.~L., {Benecchi}, S.~D., {Buie},
  M.~W., {Trilling}, D.~E., \& {Wasserman}, L.~H. 2014, \aj, 148, 55

\bibitem[{Alexandersen(2015)}]{mikephd}
Alexandersen, M. 2015, PhD thesis, University of British Columbia

\bibitem[{{Alexandersen} {et~al.}(2014){Alexandersen}, {Gladman}, {Kavelaars},
  {Petit}, {Gwyn}, \& {Shankman}}]{Alexandersen2015}
{Alexandersen}, M., {Gladman}, B., {Kavelaars}, J.~J., {Petit}, J.-M., {Gwyn},
  S., \& {Shankman}, C. 2014, ArXiv e-prints

\bibitem[{{Bannister} {et~al.}(2016){Bannister}, {Kavelaars}, {Petit},
  {Gladman}, {Gwyn}, {Chen}, {Volk}, {Alexandersen}, {Benecchi}, {Delsanti},
  {Fraser}, {Granvik}, {Grundy}, {Guilbert-Lepoutre}, {Hestroffer}, {Ip},
  {Jakubik}, {Jones}, {Kaib}, {Lacerda}, {Lawler}, {Lehner}, {Lin}, {Lister},
  {Lykawka}, {Monty}, {Marsset}, {Murray-Clay}, {Noll}, {Parker}, {Pike},
  {Rousselot}, {Rusk}, {Schwamb}, {Shankman}, {Sicardy}, {Vernazza}, \&
  {Wang}}]{Bannister2015}
{Bannister}, M.~T., {et~al.} 2016, ArXiv e-prints

\bibitem[{{Batygin} {et~al.}(2011){Batygin}, {Brown}, \&
  {Fraser}}]{Batygin2011}
{Batygin}, K., {Brown}, M.~E., \& {Fraser}, W.~C. 2011, \apj, 738, 13

\bibitem[{{Bernstein} \& {Khushalani}(2000)}]{Bernstein2000}
{Bernstein}, G., \& {Khushalani}, B. 2000, \aj, 120, 3323

\bibitem[{{Brown}(2001)}]{Brown2001}
{Brown}, M.~E. 2001, \aj, 121, 2804

\bibitem[{{Chiang} \& {Jordan}(2002)}]{Chiang2002}
{Chiang}, E.~I., \& {Jordan}, A.~B. 2002, \aj, 124, 3430

\bibitem[{{Chiang} {et~al.}(2003){Chiang}, {Jordan}, {Millis}, {Buie},
  {Wasserman}, {Elliot}, {Kern}, {Trilling}, {Meech}, \& {Wagner}}]{Chiang2003}
{Chiang}, E.~I., {et~al.} 2003, \aj, 126, 430

\bibitem[{{Dawson} \& {Murray-Clay}(2012)}]{Dawson2012}
{Dawson}, R.~I., \& {Murray-Clay}, R. 2012, \apj, 750, 43

\bibitem[{{Elliot} {et~al.}(2005){Elliot}, {Kern}, {Clancy}, {Gulbis},
  {Millis}, {Buie}, {Wasserman}, {Chiang}, {Jordan}, {Trilling}, \&
  {Meech}}]{Elliot2005}
{Elliot}, J.~L., {et~al.} 2005, \aj, 129, 1117

\bibitem[{{Fernandez} \& {Ip}(1984)}]{Fernandez1984}
{Fernandez}, J.~A., \& {Ip}, W.-H. 1984, \icarus, 58, 109

\bibitem[{{Fraser} {et~al.}(2014){Fraser}, {Brown}, {Morbidelli}, {Parker}, \&
  {Batygin}}]{Fraser2014}
{Fraser}, W.~C., {Brown}, M.~E., {Morbidelli}, A., {Parker}, A., \& {Batygin},
  K. 2014, \apj, 782, 100

\bibitem[{{Fraser} \& {Kavelaars}(2009)}]{Fraser2009}
{Fraser}, W.~C., \& {Kavelaars}, J.~J. 2009, \aj, 137, 72

\bibitem[{{Fraser} {et~al.}(2015){Fraser}, {Bannister}, {Pike}, {Schwamb},
  {Marrset}, {Kavelaars}, {Benecchi}, {Delsanti}, {Guilbert-Lepoutre},
  {Parker}, {Peixinho}, {Vernazza}, \& {Wang}}]{Fraser2015}
{Fraser}, W.~C., {et~al.} 2015, in AAS/Division for Planetary Sciences Meeting
  Abstracts, Vol.~47, AAS/Division for Planetary Sciences Meeting Abstracts,
  211.15

\bibitem[{{Fuentes} {et~al.}(2009){Fuentes}, {George}, \&
  {Holman}}]{Fuentes2009}
{Fuentes}, C.~I., {George}, M.~R., \& {Holman}, M.~J. 2009, \apj, 696, 91

\bibitem[{{Gallardo} \& {Ferraz-Mello}(1998)}]{Gallardo1998}
{Gallardo}, T., \& {Ferraz-Mello}, S. 1998, \planss, 46, 945

\bibitem[{{Gladman} {et~al.}(2008){Gladman}, {Marsden}, \&
  {Vanlaerhoven}}]{Gladman2008}
{Gladman}, B., {Marsden}, B.~G., \& {Vanlaerhoven}, C. 2008, {Nomenclature in
  the Outer Solar System}, ed. M.~A. {Barucci}, H.~{Boehnhardt}, D.~P.
  {Cruikshank}, A.~{Morbidelli}, \& R.~{Dotson}, 43--57

\bibitem[{{Gladman} {et~al.}(2012){Gladman}, {Lawler}, {Petit}, {Kavelaars},
  {Jones}, {Parker}, {Van Laerhoven}, {Nicholson}, {Rousselot}, {Bieryla}, \&
  {Ashby}}]{Gladman2012}
{Gladman}, B., {et~al.} 2012, \aj, 144, 23

\bibitem[{{Gulbis} {et~al.}(2010){Gulbis}, {Elliot}, {Adams}, {Benecchi},
  {Buie}, {Trilling}, \& {Wasserman}}]{Gulbis2010}
{Gulbis}, A., {Elliot}, J., {Adams}, E., {Benecchi}, S., {Buie}, M.,
  {Trilling}, D., \& {Wasserman}, L. 2010, \aj, in press

\bibitem[{{Hahn} \& {Malhotra}(2005)}]{Hahn2005}
{Hahn}, J.~M., \& {Malhotra}, R. 2005, \aj, 130, 2392

\bibitem[{{Johansen} {et~al.}(2007){Johansen}, {Oishi}, {Mac Low}, {Klahr},
  {Henning}, \& {Youdin}}]{Johansen2007}
{Johansen}, A., {Oishi}, J.~S., {Mac Low}, M.-M., {Klahr}, H., {Henning}, T.,
  \& {Youdin}, A. 2007, \nat, 448, 1022

\bibitem[{{Kavelaars} {et~al.}(2009){Kavelaars}, {Jones}, {Gladman}, {Petit},
  {Parker}, {Van Laerhoven}, {Nicholson}, {Rousselot}, {Scholl}, {Mousis},
  {Marsden}, {Benavidez}, {Bieryla}, {Campo Bagatin}, {Doressoundiram},
  {Margot}, {Murray}, \& {Veillet}}]{Kavelaars2009}
{Kavelaars}, J.~J., {et~al.} 2009, \aj, 137, 4917

\bibitem[{{Lawler} \& {Gladman}(2013)}]{Lawler2013}
{Lawler}, S.~M., \& {Gladman}, B. 2013, \aj, 146, 6

\bibitem[{{Levison} \& {Duncan}(1994)}]{Levison1994}
{Levison}, H.~F., \& {Duncan}, M.~J. 1994, \icarus, 108, 18

\bibitem[{{Levison} {et~al.}(2008){Levison}, {Morbidelli}, {Vanlaerhoven},
  {Gomes}, \& {Tsiganis}}]{Levison2008}
{Levison}, H.~F., {Morbidelli}, A., {Vanlaerhoven}, C., {Gomes}, R., \&
  {Tsiganis}, K. 2008, \icarus, 196, 258

\bibitem[{{Lykawka} \& {Mukai}(2007{\natexlab{a}})}]{Lykawka2007}
{Lykawka}, P.~S., \& {Mukai}, T. 2007{\natexlab{a}}, Icarus, 189, 213

\bibitem[{{Lykawka} \& {Mukai}(2007{\natexlab{b}})}]{Lykawka2007s}
---. 2007{\natexlab{b}}, \icarus, 192, 238

\bibitem[{{Malhotra}(1993)}]{Malhotra1993}
{Malhotra}, R. 1993, \nat, 365, 819

\bibitem[{{Malhotra}(1995)}]{Malhotra1995}
---. 1995, \aj, 110, 420

\bibitem[{{Morbidelli} {et~al.}(2008){Morbidelli}, {Levison}, \&
  {Gomes}}]{Morbidelli2008}
{Morbidelli}, A., {Levison}, H.~F., \& {Gomes}, R. 2008, {The Dynamical
  Structure of the Kuiper Belt and Its Primordial Origin}, ed. {Barucci, M.~A.,
  Boehnhardt, H., Cruikshank, D.~P., Morbidelli, A., \& Dotson, R.}, 275--292

\bibitem[{{Morbidelli} {et~al.}(1995){Morbidelli}, {Thomas}, \&
  {Moons}}]{Morbidelli1995}
{Morbidelli}, A., {Thomas}, F., \& {Moons}, M. 1995, \icarus, 118, 322

\bibitem[{{Murray} \& {Dermott}(1999)}]{MurrayDermott1999}
{Murray}, C.~D., \& {Dermott}, S.~F. 1999, {Solar system dynamics} (Cambridge:
  University Press)

\bibitem[{{Murray-Clay} \& {Chiang}(2005)}]{Murray-Clay2005}
{Murray-Clay}, R.~A., \& {Chiang}, E.~I. 2005, \apj, 619, 623

\bibitem[{{Murray-Clay} \& {Chiang}(2006)}]{Murray-Clay2006}
---. 2006, \apj, 651, 1194

\bibitem[{{Nesvorny}(2015)}]{Nesvorny2015}
{Nesvorny}, D. 2015, ArXiv e-prints

\bibitem[{{Nesvorn{\'y}} \& {Roig}(2000)}]{Nesvorny2000}
{Nesvorn{\'y}}, D., \& {Roig}, F. 2000, \icarus, 148, 282

\bibitem[{{Nesvorn{\'y}} \& {Roig}(2001)}]{Nesvorny2001}
---. 2001, \icarus, 150, 104

\bibitem[{{Parker}(2015)}]{Parker2015}
{Parker}, A.~H. 2015, \icarus, 247, 112

\bibitem[{{Petit} {et~al.}(2011){Petit}, {Kavelaars}, {Gladman}, {Jones},
  {Parker}, {Van Laerhoven}, {Nicholson}, {Mars}, {Rousselot}, {Mousis},
  {Marsden}, {Bieryla}, {Taylor}, {Ashby}, {Benavidez}, {Campo Bagatin}, \&
  {Bernabeu}}]{Petit2011}
{Petit}, J.-M., {et~al.} 2011, \aj, 142, 131

\bibitem[{{Pike} {et~al.}(2015){Pike}, {Kavelaars}, {Petit}, {Gladman},
  {Alexandersen}, {Volk}, \& {Shankman}}]{Pike2015}
{Pike}, R.~E., {Kavelaars}, J.~J., {Petit}, J.~M., {Gladman}, B.~J.,
  {Alexandersen}, M., {Volk}, K., \& {Shankman}, C.~J. 2015, \aj, 149, 202

\bibitem[{{Press} {et~al.}(1992){Press}, {Teukolsky}, {Vetterling}, \&
  {Flannery}}]{Press1992}
{Press}, W.~H., {Teukolsky}, S.~A., {Vetterling}, W.~T., \& {Flannery}, B.~P.
  1992, {Numerical recipes in FORTRAN. The art of scientific computing}, 2nd
  edn. (Cambridge: University Press)

\bibitem[{{Shankman} {et~al.}(2013){Shankman}, {Gladman}, {Kaib}, {Kavelaars},
  \& {Petit}}]{Shankman2013}
{Shankman}, C., {Gladman}, B.~J., {Kaib}, N., {Kavelaars}, J.~J., \& {Petit},
  J.~M. 2013, \apjl, 764, L2

\bibitem[{{Shankman} {et~al.}(2016){Shankman}, {Kavelaars}, {Gladman},
  {Alexandersen}, {Kaib}, {Petit}, {Bannister}, {Chen}, {Gwyn}, {Jakubik}, \&
  {Volk}}]{Shankman2016}
{Shankman}, C., {et~al.} 2016, \aj, 151, 31

\bibitem[{{Sheppard}(2012)}]{Sheppard2012}
{Sheppard}, S.~S. 2012, \aj, 144, 169

\bibitem[{{Thomas} \& {Morbidelli}(1996)}]{Thomas1996}
{Thomas}, F., \& {Morbidelli}, A. 1996, Celestial Mechanics and Dynamical
  Astronomy, 64, 209

\bibitem[{{Thommes} {et~al.}(1999){Thommes}, {Duncan}, \&
  {Levison}}]{Thommes1999}
{Thommes}, E.~W., {Duncan}, M.~J., \& {Levison}, H.~F. 1999, \nat, 402, 635

\bibitem[{{Tiscareno} \& {Malhotra}(2009)}]{Tiscareno2009}
{Tiscareno}, M.~S., \& {Malhotra}, R. 2009, \aj, 138, 827

\bibitem[{{Tsiganis} {et~al.}(2005){Tsiganis}, {Gomes}, {Morbidelli}, \&
  {Levison}}]{Tsiganis2005}
{Tsiganis}, K., {Gomes}, R., {Morbidelli}, A., \& {Levison}, H.~F. 2005, \nat,
  435, 459

\bibitem[{{Volk} \& {Malhotra}(2011)}]{Volk2011}
{Volk}, K., \& {Malhotra}, R. 2011, \apj, 736, 11

\bibitem[{{Wan} \& {Huang}(2007)}]{Wan2007}
{Wan}, X.-S., \& {Huang}, T.-Y. 2007, \mnras, 377, 133

\bibitem[{{Wolff} {et~al.}(2012){Wolff}, {Dawson}, \&
  {Murray-Clay}}]{Wolff2012}
{Wolff}, S., {Dawson}, R.~I., \& {Murray-Clay}, R.~A. 2012, \apj, 746, 171

\bibitem[{{Youdin} \& {Goodman}(2005)}]{Youdin2005}
{Youdin}, A.~N., \& {Goodman}, J. 2005, \apj, 620, 459

\bibitem[{{Yu} \& {Tremaine}(1999)}]{Yu1999}
{Yu}, Q., \& {Tremaine}, S. 1999, \aj, 118, 1873

\end{thebibliography}
\end{document}